\documentclass[twocolumn,amsmath,amssymb,showpacs,pre,superscriptaddress]{revtex4}
\usepackage[latin1]{inputenc}
\usepackage[dvips]{epsfig}

\bibstyle{apsrev}
\begin{document}

\title[Structural transitions in sheared binary immiscible fluids and
microemulsions]{Structural transitions and arrest of domain growth in
sheared binary immiscible fluids and microemulsions} 

\author{J. Harting}
\affiliation{Institute for Computational Physics, University of Stuttgart,
Pfaffenwaldring 27, 70569 Stuttgart, Germany }
\author{G. Giupponi}
\affiliation{Centre for Computational Science, Department of Chemistry,
University College London, 20 Gordon Street, WC1H 0AJ, London, UK}
\author{P.V. Coveney}
\affiliation{Centre for Computational Science, Department of Chemistry,
University College London, 20 Gordon Street, WC1H 0AJ, London, UK}

\date{\today}

\newcommand{\eq}{Eq.\,}
\newcommand{\tab}{Tab.\,}
\newcommand{\fig}{Fig.\,}
\newcommand{\figs}{Figs.\,}

\begin{abstract}
We investigate spinodal decomposition and structuring effects in binary
immiscible and ternary amphiphilic fluid mixtures under shear by means of three
dimensional lattice Boltzmann simulations. We show that the growth of
individual fluid domains can be arrested by adding surfactant to the system,
thus forming a bicontinuous microemulsion. We demonstrate that the maximum
domain size and the time of arrest depend linearly on the concentration of
amphiphile molecules. In addition, we find that for a well defined threshold
value of amphiphile concentration, the maximum domain size and time of complete
arrest do not change. For systems under constant and oscillatory shear we
analyze domain growth rates in directions parallel and perpendicular to the
applied shear. We find a structural transition from a sponge to a lamellar
phase by applying a constant shear and the occurrence of tubular structures
under oscillatory shear. The size of the resulting lamellae and tubes depends
strongly on the amphiphile concentration, shear rate and shear frequency.
\end{abstract}

\pacs{64.75.+g,47.11.-j, 82.70.Uv, 83.60.-a}
%\pacs{47.11.-j}{Computational methods in fluid dynamics}
%\pacs{82.70.Uv}{Surfactants, micellar solutions, vesicles, lamellae,
%  amphiphilic systems, (hydrophilic and hydrophobic interactions)}
%\pacs{83.60.-a}{Rheology, Material behaviour}
\maketitle

\section{Introduction}
Complex fluid mixtures consisting of immiscible fluid species and
surfactants are ubiquitous in many industrial applications. These fluid
mixtures involve both hydrodynamic flow effects and complex interactions
between fluid particles. Typical examples can be found in the food,
cosmetic and chemical industries where surfactants are utilized to stabilize
otherwise immiscible fluids. A good example is a barbecue sauce containing
large fractions of water and oil or fat. Without any additives the constituents
would phase separate, entering a not very appealing de-mixed state in which the
fat accumulates in a thick layer on top of the remainder. Adding an emulsifier
or surfactant helps to stabilize the sauce. These molecules are often called
amphiphiles and, in their simplest form, are comprised of a hydrophilic
(water-loving) head group and a hydrophobic (oil-loving) tail.  The surfactant
molecules self-assemble on the surface of oil droplets and reduce the surface
tension. Thus, the droplets stabilize and remain suspended within the bulk
water. A typical emulsifier used by the food industry is egg yolk lecithin.
Proteins and emulsifiers with low molecular weight are also common emulsifiers.

Amphiphilic fluids containing at least one species made of surfactant
molecules are in general complex fluids that can self-assemble to form
ordered structures such as lamellae, micelles, not ordered sponge phases
and liquid crystalline (cubic) phases. In general, adding amphiphiles to
a binary mixture of otherwise immiscible fluids (for example oil and
water) which is undergoing phase separation can cause the de-mixing
process to slow down. If the amphiphile concentration is sufficiently
high, the de-mixing process might eventually arrest completely. It has been
shown by Langevin, molecular dynamics, lattice gas, and lattice Boltzmann
simulations that the temporal growth law for the size of oil and water
domains in a system without amphiphiles follows a power law
$t^\alpha$~\cite{bib:bray,bib:gonzalez-nekovee-coveney} and crosses over
to a logarithmic growth law $(\ln t)^\theta$, where $\alpha$, $\theta$ are
fitting parameters and $t$ is the
time~\cite{bib:kawakatsu-kawasaki-1993,bib:emerton-coveney-boghosian,bib:gonzalez-coveney-2}.
A further increase of the surfactant concentration can lead to growth
which is well described by a stretched exponential form $A-B\exp(-C t^D)$,
where capital letters denote fitting
parameters~\cite{bib:emerton-coveney-boghosian,bib:gonzalez-coveney-2}.
By adjusting temperature, fluid composition or pressure, amphiphiles can
self-assemble and force the fluid mixture into a number of equilibrium
structures or mesophases. These include lamellae and hexagonally packed
cylinders, micellar, primitive, diamond, or gyroid cubic mesophases as
well as sponge phases. In this paper we focus on the sponge mesophase,
which in the context of our simulations is called a bicontinuous
microemulsion since it is formed by the amphiphilic stabilization of a
phase-separating binary mixture, where the immiscible fluid constituents
occur in equal proportions. Here, the oil and water phases interpenetrate
and percolate and are separated by a monolayer of surfactant at the
interface.

Such complex fluids are often subject to shear forces and show pronounced
rheological properties~\cite{bib:JonesBook,bib:gompper-schick}.  Often,
shear induced transitions from isotropic to lamellar phases can be
observed. These have been studied experimentally in
binary~\cite{bib:meyer-2000} and ternary amphiphilic
fluids~\cite{bib:zipfel-1999,bib:berghausen-2000}. If oscillatory shear is
applied, a further transition to a tubular phase or a transition between
differently oriented lamellar phases can
occur~\cite{bib:qiu-zhang-yang:1998,bib:zhang-wiesner-yang-pakula-spiess:1996,bib:zhang-wiesner:1995,bib:xu-gonnella-lamura:2003}.

Computationally, such complex fluids are too large and expensive to
tackle with atomistic methods such as molecular dynamics, yet they require
too much molecular detail for continuum Navier-Stokes approaches.
Algorithms which work at an intermediate or ``mesoscale'' level of
description have been developed during the last twenty years, including
dissipative particle
dynamics\cite{bib:espanol-warren,bib:jury-bladon-cates,bib:flekkoy-coveney-defabritiis},
lattice gas cellular automata\cite{bib:rivet-boon}, the stochastic
rotation dynamics \cite{bib:malevanets-kapral,bib:hashimoto-chen-ohashi,bib:sakai-chen-ohashi},
and the lattice Boltzmann
equation\cite{bib:succi,bib:benzi-succi-vergassola,bib:love-nekovee-coveney-chin-gonzalez-martin}.
In particular, the lattice Boltzmann method has been found highly useful
for simulating complex fluid flows in a wide variety of systems. This
algorithm, described in more detail in the next section, is extremely well
suited to implementation on parallel computers, which permits very large
systems to be simulated, reaching hitherto inaccessible physical regimes.

In this paper we investigate spinodal decomposition and structuring
effects in binary immiscible and ternary amphiphilic fluid mixtures under
shear by means of large scale three dimensional lattice Boltzmann (LB)
simulations. The purely kinetic LB method we use is able to model complex
flows whose rheological properties are emergent from the mesoscopic
kinetic processes without any imposed macroscopic
constraints\cite{bib:chen-boghosian-coveney}.

Varieties of the lattice Boltzmann method have been used successfully to
study the behaviour of multi-phase flows in the past.  A number of authors
have investigated spinodal
decomposition~\cite{bib:osborn-orlandini-swift-yeomans-banavar,bib:chin-coveney,bib:gonzalez-nekovee-coveney,bib:chin-boek-coveney,bib:alexander-chen-grunau,bib:wagner-yeomans,bib:gonnella-orlandini-yeomans,bib:cates-kendon-bladon-desplat,bib:kendon-desplat-bladon-cates,bib:kendon-cates-pagonabarraga-desplat-bladon,bib:pagonabarraga-desplat-wagner-cates:2001,bib:xu-gonnella-lamura:2004}
and the same phenomenon has also attracted some interest in the presence of
shear, where structural transitions from isotropic to lamellar or
tubular phases may
occur~\cite{bib:wagner-yeomans-shear,bib:wagner-pagonabarraga,bib:jens-venturoli-coveney:2004,bib:stansell-stratford-desplat-adhikari-cates:2006,bib:xu-gonnella-lamura:2003}.

There have been only limited investigations of
 the influence of amphiphiles on the domain growth within the lattice
Boltzmann method, despite the fact that ternary amphiphilic fluids have
been studied by a number of
authors~\cite{bib:theissen-gompper-kroll,bib:lamura-gonnella-yeomans,bib:nekovee-coveney-chen-boghosian,bib:love-nekovee-coveney-chin-gonzalez-martin}.
For example, it has been shown that the lattice Boltzmann method can be
used to describe the self-assembly and the rheological properties of
mesophases including the primitive $P$-phase~\cite{bib:nekovee-coveney}
and the gyroid
phase\cite{bib:jens-giupponi-coveney:2006}. The gyroid mesophase in particular has been of major interest during
the last few years, where the phase
formation and structural properties
\cite{bib:gonzalez-coveney,bib:gonzalez-coveney-2}, the influence of
defects
\cite{bib:jens-teragyroid:2004,bib:jens-harvey-chin-coveney:2004,bib:jens-harvey-chin-venturoli-coveney:2005,bib:chin-coveney06},
as well as its properties under shear
\cite{bib:jens-giupponi-coveney:2006,bib:jens-gonzalez-giupponi-coveney:2005}
have been investigated.
%Chin and Coveney\cite{bib:chin-coveney06} recently discovered that the self assembly of the
%G (gyroid) mesophase starting from a random ternary amphiphilic mixture is
%accompanied by the formation of multiple domains separated by walls, which
%eventually disappear merging one another. 

In this paper, our aim is to focus on the effect of surfactant
concentration on the length and time scales of arrested growth and on the
changes in structural property induced by steady or oscillatory shear.

The remainder of the paper is organized as follows. After an introduction
to the simulation method and the fluid parameters used, we present results from
the simulation of ternary amphiphilic systems with varying surfactant
density. Here, we draw on the results of Gonz\'{a}lez-Segredo et
al.~\cite{bib:gonzalez-coveney-2} for the simulation parameters used
within the present work. Thereafter, we extend our study to systems under
constant and oscillatory shear, where we report on the influence of the
domain growth rates and structural transitions induced by the shear.
Finally, a summary and conclusions are given.
%We supplement our study with visualization evidence in the form of animations
%produced by simulation.

\section{Simulation method}
A set of equations can be used to represent a standard LB system involving multiple species~\cite{bib:higuera-succi-benzi}
\begin{equation}
\label{LBeqs}
n_i^{\alpha}({\bf x}+{\bf c}_i, t+1) - n_i^{\alpha}({\bf x},t) = 
\Omega_i^{\alpha}
\end{equation}
with $i= 0,1,\dots,b$.
The single-particle distribution function $n_i^{\alpha}({\bf x},t)$ indicates the density of species $\alpha$, having velocity ${\bf c}_i$, at site ${\bf x}$ on a
D-dimensional lattice of coordination number $b$, at timestep $t$. The
collision operator 
\begin{equation}
\Omega_i^{\alpha}=-\frac{1}{\tau^{\alpha}}(n_i^{\alpha}({\bf x},t) - n_i^{\alpha
 eq}({\bf x},t)),
\end{equation}
represents the change in the
single-particle distribution function due to the collisions. A popular
form is the single relaxation time $\tau_{\alpha}$, linear `BGK'
form~\cite{bib:Bhatnagar54} for the collision operator.  It can be shown
for low Mach numbers that the LB equations correspond to a solution of the
Navier-Stokes equation for isothermal, quasi-incompressible fluid flow.
The lattice Boltzmann method is an excellent candidate to exploit the
possibilities of parallel computers, as the dynamics at a lattice site requires
only information about quantities at nearest neighbour lattice
sites~\cite{bib:love-nekovee-coveney-chin-gonzalez-martin,bib:jens-harvey-chin-venturoli-coveney:2005}. The
local equilibrium distribution $n_i^{\alpha eq}$ plays a fundamental role
in the dynamics of the system as shown by Eq.~(\ref{LBeqs}). In this
study, we use a purely kinetic approach, for which the local equilibrium
distribution $n_i^{\alpha eq}({\bf x},t)$ is derived by imposing certain
restrictions on the microscopic processes, such as explicit mass and total
momentum conservation~\cite{bib:chen-chen-martinez-matthaeus}
\begin{equation}
\label{Equil}
\begin{array}{ll}
%n_i^{\alpha eq} = \zeta_i\rho^{\alpha}\left[1+
%\frac{{\bf c}_i \cdot {\bf u}}{c_s^2} +\frac{({\bf c}_i \cdot {\bf u})^2}{2c_s^4}
%-\frac{u^2}{2c_s^2}+\frac{({\bf c}_i \cdot {\bf u})^3}{6c_s^6}
%-\frac{u^2({\bf c}_i \cdot {\bf u})}{2c_s^4}\right] \mbox{ ,}
n_i^{\alpha eq} = \\
 \zeta_i\rho^{\alpha}\left[1+
\frac{{\bf c}_i {\bf u}}{c_s^2} +\frac{({\bf c}_i {\bf u})^2}{2c_s^4}
-\frac{u^2}{2c_s^2}+\frac{({\bf c}_i {\bf u})^3}{6c_s^6}
-\frac{u^2({\bf c}_i {\bf u})}{2c_s^4}\right],
\end{array}
\end{equation}
where 
$\rho^{\alpha}({\bf x},t)\equiv\sum_i \eta_i^{\alpha}({\bf x},t)$ is the
fluid density and
${\bf u} = {\bf u}({\bf x},t)$ is the macroscopic bulk velocity
of the fluid, given by $\rho^{\alpha}({\bf x},t){\bf u}^{\alpha}
\equiv \sum_i n_i^{\alpha}({\bf x},t){\bf c}_i$. $\zeta_i$ are the
coefficients resulting from the velocity space discretization and
$c_s$ is the speed of sound, both of which are determined by the
choice of the lattice. We use a D3Q19 implementation, i.e., a three dimensional lattice with 19 discrete velocities.
Immiscibility of species $\alpha$ is introduced in the model following
Shan and Chen~\cite{bib:shan-chen,bib:shan-chen-liq-gas}, where only nearest neighbour
interactions among the species are considered.  These
interactions are described by a self-consistently generated mean field
body force
\begin{equation}
\label{Eq:SCforce}
{\bf F}^{\alpha}({\bf x},t) \equiv -\psi^{\alpha}({\bf x},t)\sum_{\bf \bar{\alpha}}g_{\alpha \bar{\alpha}}\sum_{\bf x^{\prime}}\psi^{\bar{\alpha}}({\bf x^{\prime}},t)({\bf x^{\prime}}-{\bf x})\mbox{ ,}
\end{equation}
where $\psi^{\alpha}({\bf x},t)$ is the so-called effective mass, which
can have a general form for modeling various types of fluids (we use
$\psi^{\alpha} = (1 - e^{-\rho^{\alpha}}$)\cite{bib:shan-chen}), and
$g_{\alpha\bar{\alpha}}$ is a force coupling constant whose magnitude
controls the strength of the interaction between components $\alpha$,
$\bar{\alpha}$ and is set positive to mimic repulsion. The dynamical
effect of the force is realized in the BGK collision operator by adding to
the velocity ${\bf u}$ in the equilibrium distribution of
eq.~(\ref{Equil}) the increment
\begin{equation}
\delta{\bf u}^{\alpha} = \frac{\tau^{\alpha}{\bf F}^{\alpha}}{\rho^{\alpha}}\mbox{ .}
\end{equation}

Amphiphiles are introduced within the model as described in
~\cite{bib:chen-boghosian-coveney}
and~\cite{bib:nekovee-coveney-chen-boghosian}.  An
amphiphile usually possesses two different fragments, each having an
affinity for one of the two immiscible components. The orientation of any
amphiphile present at a lattice site ${\bf x}$ is represented by an
average dipole vector ${\bf d}({\bf x},t)$.  Its direction is allowed to
vary continuously and no information is specified for each velocity ${\bf
c}_i$, for reasons of computational efficiency and simplicity. The
amphiphile density at a given site is given by an additional fluid species
with density $\rho^s$. 
The model has been used successfully to study spinodal
decomposition~\cite{bib:chin-coveney,bib:gonzalez-nekovee-coveney}, the formation of
mesophases~\cite{Maziar:2001,bib:nekovee-coveney,bib:gonzalez-coveney,bib:gonzalez-coveney-2,bib:jens-harvey-chin-coveney:2004,bib:jens-giupponi-coveney:2006,bib:jens-gonzalez-giupponi-coveney:2005},
and flow in porous media~\cite{bib:jens-venturoli-coveney:2004}.
%
%\fixme{More on surfactants?}

We use LB3D\cite{bib:jens-harvey-chin-venturoli-coveney:2005}, a highly
scalable parallel LB code, to implement the model. The very good scaling
of our code permits us to run all our simulations on multiprocessor
machines and computational grids in order to reduce the length of data
collection to a few weeks. Also, we are able to use simulation boxes
typically eight times bigger than previous studies so as to
minimize finite size effects.

In order to study the rheological behaviour of complex fluid mixtures,
we have implemented Lees-Edwards boundary conditions, which were
originally developed for molecular dynamics simulations\cite{bib:lees-edwards}. They
reduce finite size effects as compared to moving solid
walls~\cite{bib:lees-edwards} and have been used in lattice Boltzmann
simulations by various authors
~\cite{bib:wagner-yeomans-shear,bib:wagner-pagonabarraga,bib:harting-venturoli-coveney,bib:stansell-stratford-desplat-adhikari-cates:2006}.
This computationally convenient method imposes new positions and
velocities on particles leaving the simulation box in the direction
perpendicular to the imposed shear strain while leaving the other
coordinates unchanged. Choosing $z$ as the direction of shear and $x$ as
the direction of the velocity gradient, we have
\begin{eqnarray}%\hspace{-1cm}
z^{\prime}& \equiv & \left\{
\begin{array}{ll}
(z+\Delta_z) \mbox{ mod $N_z$}&,x > N_x  \\
z \mbox{ mod $N_z$}&,0\le x \le N_x   \\
(z-\Delta_z) \mbox{ mod $N_z$}&,x < 0    \\
\end{array} \right.
\\
%&\quad
u_z^{\prime}& \equiv & \left\{
\begin{array}{ll}
u_z+U&,x > N_x  \\
u_z&,0\le x \le N_x   \\
u_z-U&,x < 0    \\
\end{array} \right.\mbox{ ,}
\end{eqnarray}
where $\Delta_z \equiv U\Delta t$, U is the shear velocity, $u_z$ is the
$z-$component of ${\bf u}$ and $N_{x(z)}$ is the system length in the
$x(z)$ direction. We also use an interpolation scheme suggested by Wagner
and Pagonabarraga~\cite{bib:wagner-pagonabarraga} as $\Delta_z$ is not
generally a multiple of the nearest neighbour lattice distance. For oscillatory shear, we set
\begin{equation}
\label{coseno}
U(t) = U\cos(\omega t)\mbox{ ,}
\end{equation}
where $\omega/2\pi$ is the frequency of oscillation. 

%This results in a maximum
%shear rate $\dot{\gamma}_{xz}=\frac{2\times0.1}{64}=3.2\times10^{-3}$
%in lattice units. 

In non-sheared studies of spinodal
decomposition it has been shown that large lattices are needed to overcome
finite size effects.  There, 128$^3$ was the minimum acceptable number of
lattice sites \cite{bib:gonzalez-nekovee-coveney}. More quantitatively, Kendon
et al.\cite{bib:kendon-desplat-bladon-cates} set $L/4$ as the maximum length scale which is not
affected by finite size effects in their spinodal decomposition simulation, 
where $L$ is the length of the simulation box.
We therefore choose
256$^3$ for all non-sheared simulations to limit the influence of finite size
effects even further. For high shear
rates, systems also have to be very extended in the direction of the applied shear because, if the system is too
small, the domains interconnect across the $z = 0$ and $z = N_z$
boundaries to form continuous lamellae in the direction of
shear~\cite{bib:jens-venturoli-coveney:2004,bib:jens-harvey-chin-venturoli-coveney:2005}.
Such artefacts need to be eliminated from our simulations. In this case, a good
compromise to limit finite size effects and to keep the computational
expense as low as possible is a lattice size of 128x128x512 and this is used here. Mass and
relaxation times are always set to unity, i.e. $\tau^\alpha$=1.0,
$m^\alpha$=1.0.  
We call the two immiscible fluids ``red'' and ``blue'' and set their
initial densities to identical values, $\rho^r=\rho^b$.  The initial
average surfactant density $\rho^s$ is varied between 0.0 and 0.7. The
lattice is than randomly populated with constant initial total fluid densities
$\rho^{\rm tot}=\rho^r+\rho^b+\rho^s = 1.6$. This is in
contrast to previous studies where only $\rho^r+\rho^b$ was kept
constant\cite{bib:gonzalez-coveney-2}.
The coupling constant in Eq.~\ref{Eq:SCforce} between ``red'' and ``blue'' species is set to
$g_{\rm br}$=0.08, the coupling between an immiscible fluid and surfactant
to $g_{\rm bs}$=-0.006 and the constant describing the strength of the
surfactant-surfactant interaction is kept at $g_{\rm ss}=-0.003$.
All units in this paper are given in lattice units if not stated
otherwise. While our method has been used to simulate other mesophases like
lamellar phases, the primitive $P$-phase~\cite{bib:nekovee-coveney}, and the
gyroid phase~\cite{bib:gonzalez-coveney,bib:gonzalez-coveney-2}, the
parameters used in all simulations presented here are known to produce a
sponge phase in the absence of bulk flow. More detailed investigations of
the particular choice of coupling constants and how they modify the
system's properties are given
in~\cite{bib:gonzalez-nekovee-coveney,bib:gonzalez-coveney,bib:gonzalez-coveney-2,bib:jens-gonzalez-giupponi-coveney:2005,bib:jens-giupponi-coveney:2006}.

To analyze the behaviour of the various simulations, we define the time dependent
lateral domain size $L(t)$ along direction $i=x,y,z$ as
\begin{equation}
\label{eq:domsize}
L_i(t)\equiv \frac{2\pi}{\sqrt{\left< k^2_i(t)\right>}},
\end{equation}
where
\begin{equation}
\left<k^2_i(t)\right>\equiv \frac{\sum_\mathbf{k} k_i^2 S(\mathbf{k},t)}
{\sum_\mathbf{k} S(\mathbf{k},t)} 
\end{equation}
is the second order moment of the three-dimensional structure function
\begin{equation}
S(\mathbf{k},t)\equiv\frac{1}{V}\left|\phi^\prime_\mathbf{k}(t)\right|^2
\end{equation}
with respect to the Cartesian component $i$, $\left< \right>$ denotes the
average in Fourier space, weighted by $S(\mathbf{k}, t)$ and $V$ is the
number of nodes of the lattice, $\phi^\prime_\mathbf{k}(t)$ the Fourier
transform of the fluctuations of the order parameter
$\phi^\prime\equiv\phi-\left<\phi\right>$, and $k_i$ is the $i$th
component of the wave vector. A projection of the structure function allows
us to compare simulation data to scattering patterns obtained in experiments.
We obtain those projections by summing up $S(\mathbf{k},t)$ in one of the
Cartesian directions. For example, for the projection in the $z$-direction
this leads to $S_z(k_x,k_y,t)=\sum_{k_z}S(\mathbf{k},t)$.
\section{Results}
\begin{figure*}
%\centerline{\epsfig{file=fig256cubsh0-t30000.ps,width=0.80\linewidth}}
\centerline{\epsfig{file=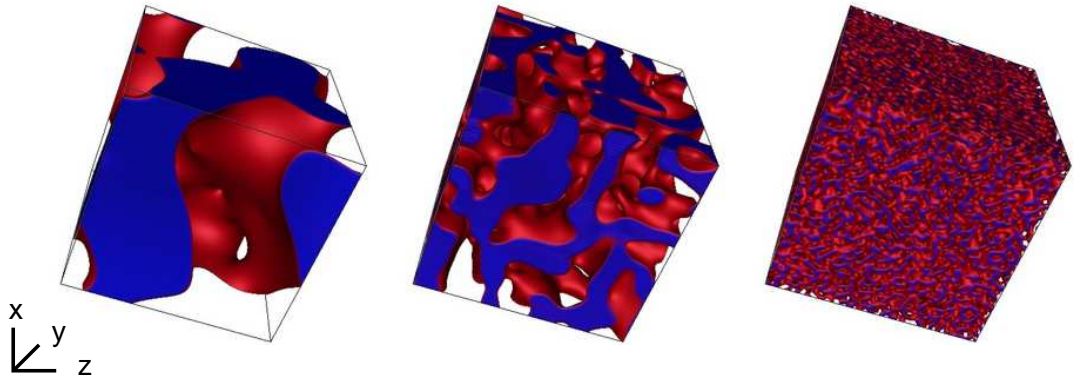,width=0.80\linewidth}}
\caption{\label{Fig:fig256cubsh0-t30000}(Color online) Volume rendered fluid densities of
256$^3$ systems at $t$=30000 for surfactant densities $\rho^s$=0.00, 0.15,
0.30 (from left to right). For better visibility only one of the
immiscible fluid species is shown. Different colours denote the interface
and areas of high density of the visualized fluid. The surfactant particles (not shown)
are aligned at the interfaces and the second immiscible fluid component fills
the void space. After 30000 timesteps the phases have separated to a large
extent if no surfactant is present (left). Adding a small amount of
surfactant ($\rho^s=0.15$, center) causes the domains to grow more slowly,
as depicted by the smaller structures in the volume rendered image. For
sufficiently high amphiphile concentrations ($\rho^s=0.30$, right) the
growth process arrests with the formation of a stable bicontinuous microemulsion.}
% \href{256cub-0-015-03.avi}{(movie available online)}.}
\end{figure*}
\subsection{Ternary amphiphilic systems without shear}
Spinodal decomposition of a binary immiscible fluid mixture has been
studied extensively within our model by Gonz\'{a}lez-Segredo et
al.~\cite{bib:gonzalez-nekovee-coveney}. The authors report domain size
scaling as expected for a crossover from diffusive behaviour to
hydrodynamic viscous growth, i.e. the domain size grows as $L\propto
t^\gamma$, with $\gamma$ being between $1/3$ and $1$. Moreover, they find
very good agreement with the dynamical scaling hypothesis, recovering the
expected universal behaviour of the structure function.

If one adds surfactant to a binary immiscible fluid mixture, the
surfactant molecules self-assemble at the interface between the two
species and slow down the phase separation process. For sufficiently high
surfactant concentrations, domain growth is arrested completely
leading to a stable microemulsion.  In~\cite{bib:gonzalez-coveney-2},
Gonz\'{a}lez-Segredo et al. extend their work to ternary amphiphilic
fluids and study how the phase separation of a binary immiscible fluid
mixture is altered by the presence of surfactant. As already described in
the introduction, by gradually increasing the surfactant concentration
they find a slowing down of the domain growth, initially from algebraic to
logarithmic temporal dependence, and, at higher surfactant concentrations,
from logarithmic to stretched-exponential behaviour. They also observe a
structural transition from sponge to gyroid phases by increasing the
amphiphile concentration or varying the amphiphile-amphiphile or
amphiphile-binary fluid coupling constants. For growth-arrested
mesophases, they observe temporal oscillations of the domain size due to
Marangoni flows.

In the present work we use simulation parameters which differ from
previous studies and which are known to produce a sponge phase. In order
to avoid effects due to variations of the fluid densities, we also keep
the total density in the system constant at 1.6 (in lattice units), while
varying the surfactant densities $\rho^s$ between 0.00 and 0.70.
Furthermore, our simulation lattices are up to eight times larger in order
to keep the influence of finite size effects to a minimum and we simulate
for up to 30000 timesteps in order to gain a better understanding of the
long time behaviour of the system.  We study the influence of the
amphiphile concentration on the phase separation process in this section
and reproduce previous results with the present parameters. In addition,
we study the dependence of the maximum achievable domain size in a stable
microemulsion on surfactant concentration as well as the
time needed to achieve this state. 

To depict the influence of the surfactant density on the phase separation
process, Fig.~\ref{Fig:fig256cubsh0-t30000} shows three volume rendered
256$^3$ systems at surfactant densities 0.0 (left), 0.15 (center), and 0.3
(right). As in all figures throughout the paper, for better visibility
only one of the immiscible fluid species is shown. Different colours
denote the interface and  areas of high density of the rendered fluid. The
surfactant particles are aligned at the interfaces and the second
immiscible constituent fills the void space. After 30000 timesteps the
phases have separated to a large extent when no surfactant is present
(left). Running the simulation for even longer would result in two
perfectly separated phases, each of them contained in a single domain
only. If one adds some surfactant ($\rho^s=0.15$, center), the domains
grow more slowly, visualized by the smaller structures in the volume
rendered image. For sufficiently high amphiphile concentrations
($\rho^s=0.30$, right) the growth process arrests leading to a stable
bicontinuous microemulsion with small individual domains formed by the two
immiscible fluids.

\begin{figure}
\centerline{a)\epsfig{file=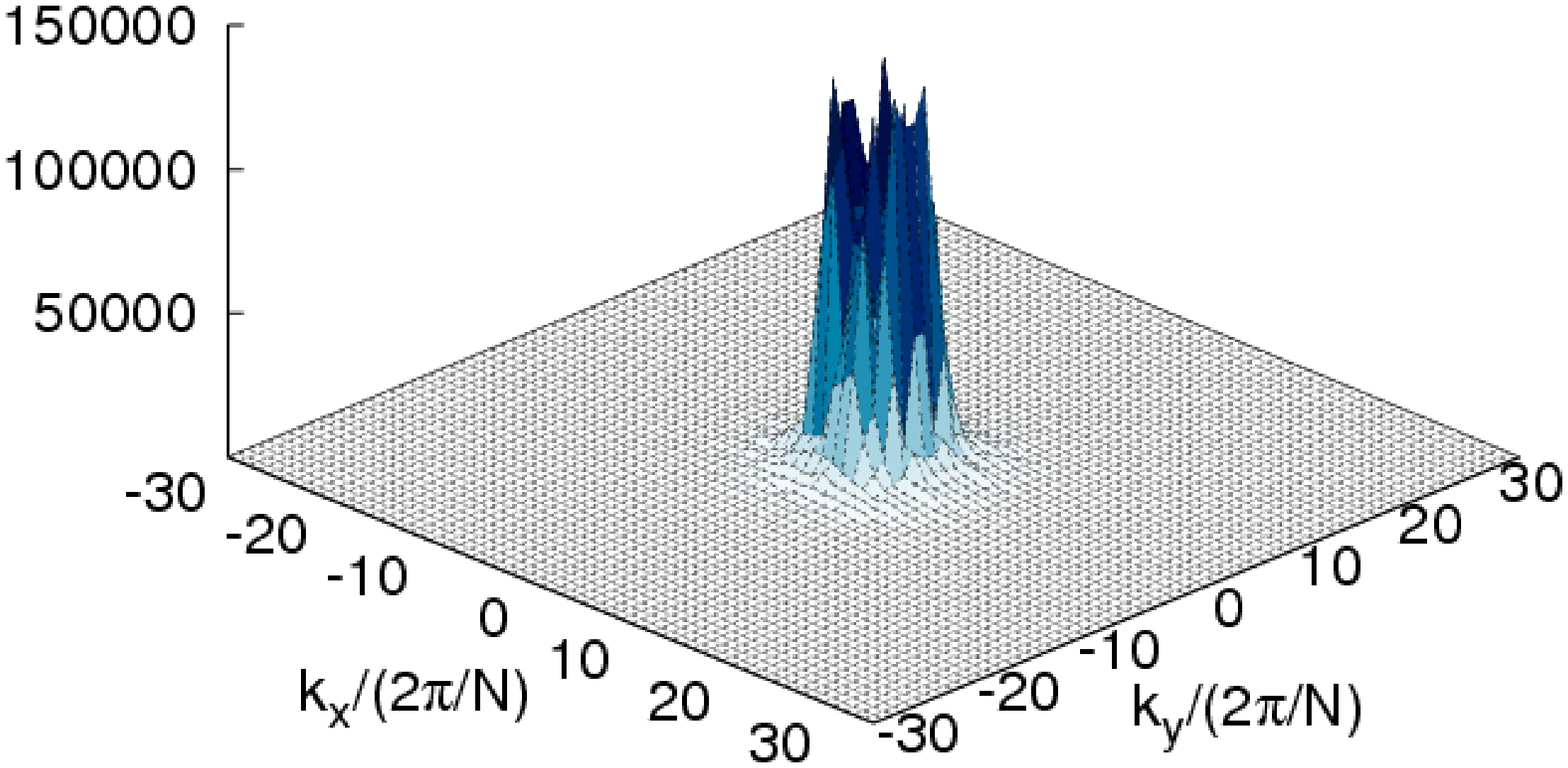,width=0.40\linewidth,angle=270}\hspace{-0.1\linewidth}
b)\epsfig{file=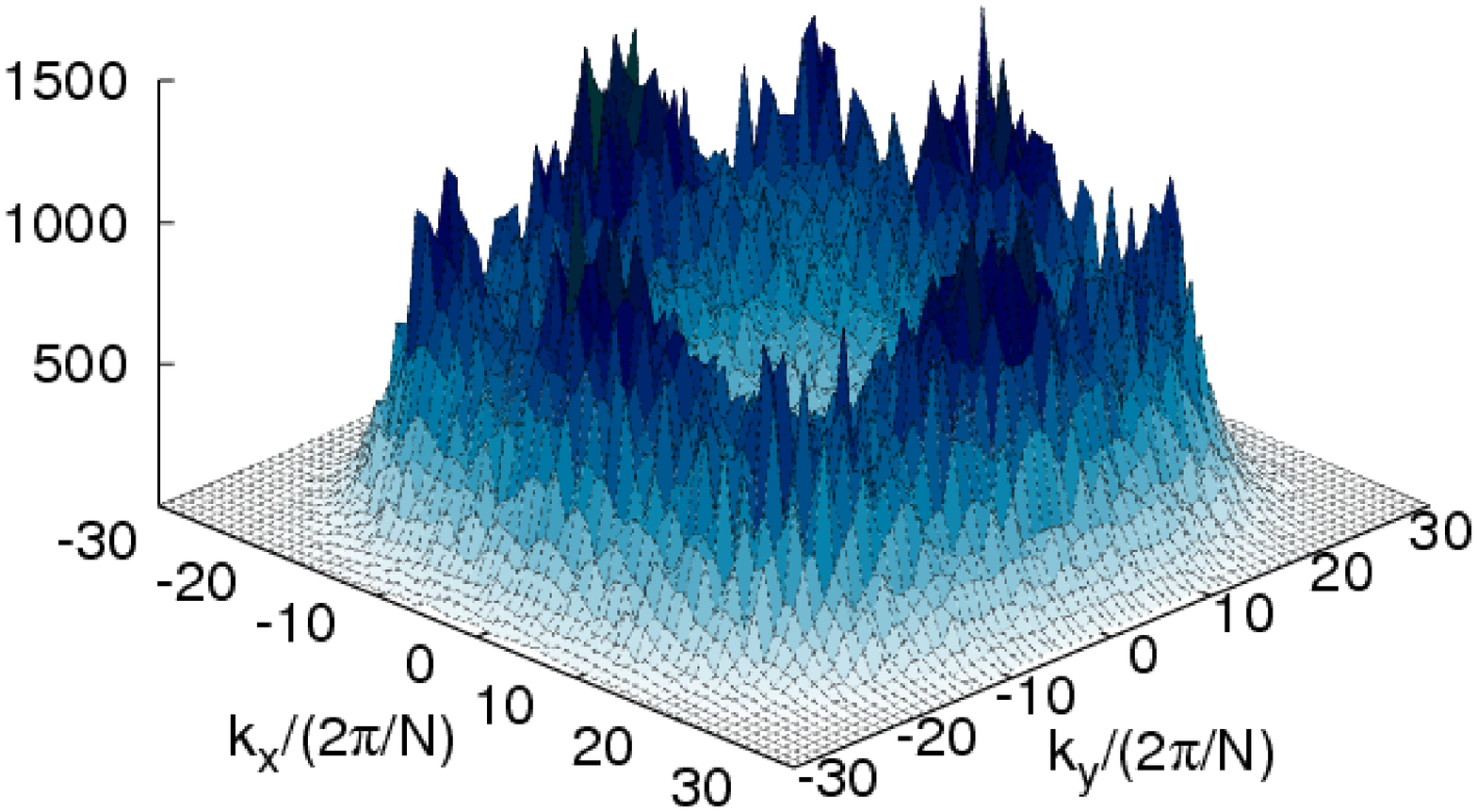,width=0.40\linewidth,angle=270}
%\centerline{a)\epsfig{file=colour_Ftot1.6_256_sh0fg0_t010000.sf3dxsum.pre.eps,width=0.40\linewidth,angle=270}\hspace{-0.1\linewidth}
%b)\epsfig{file=colour_Ftot1.6_256_sh0fg0.3_t010000.sf3dxsum.pre.eps,width=0.40\linewidth,angle=270}
}
\caption{\label{Fig:SFsh0fgvar}(Color online) Projected structure function
$S_z(k_x,k_y,t)$ for (a) $\rho^s$=0.00 and (b) 0.30 at timestep $t=10000$.
For the case without surfactant, a strong peak occurs for small values of
$k_x,k_y$ reflecting the dominance of length
scales which are of the order of the system size. For $\rho^s$=0.30, only much smaller peaks occur for larger values
of $k_x,k_y$ indicating that only small length scales are present. All
quantities are expressed in lattice units.}
\end{figure}

The projected structure function $S_z(k_x,k_y,t)$ (``scattering pattern'')
is given in Fig.~\ref{Fig:SFsh0fgvar} for two surfactant densities
$\rho^s$=0.00 (a) and 0.30 (b) at timestep $t=10000$. As can be clearly
seen in Fig.~\ref{Fig:SFsh0fgvar}a), a strong peak occurs for small values
of $k_x,k_y$ depicting the occurrence of length
scales which are of the order of the system size. For $\rho^s$=0.30, however, the peaks are by a factor of 100
smaller and shifted to larger values of $k_x,k_y$. We find a volcano-like
scattering pattern indicating the dominance of small length scales.
Since our system is cubic and no shear is applied, the projections of the
structure function in $x$ and $y$ direction are equivalent. 

\begin{figure}
%\centerline{\epsfig{file=sh0fgvar-vs-t-linlog.eps,width=0.90\linewidth}}
\centerline{\epsfig{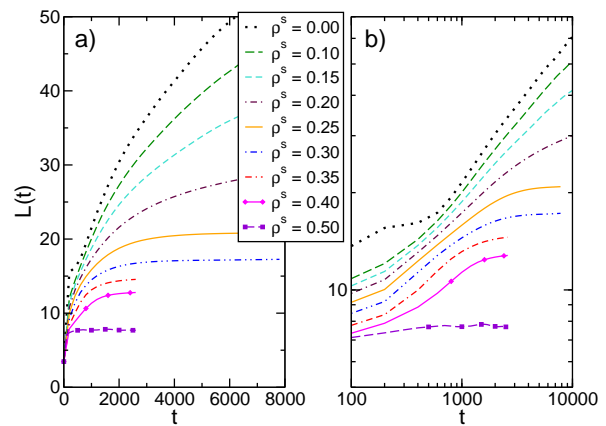}}
\caption{\label{Fig:sh0fgvar-vs-t}(Color online) Average domain size $L(t)$ for various
surfactant densities $\rho^s$=0.00, 0.10, 0.15, 0.20, 0.25, 0.30,
0.35, 0.40, 0.50. Axes are linear in (a) and logarithmic in (b). After
the initial spontaneous formation of domains, domain growth can be
described by a power law. With increasing $\rho^s$ domain growth slows
down and eventually comes to an end at a maximum domain size $L_{\rm
max}(\rho^s)$. All quantities are expressed in lattice units.}
\end{figure}

To investigate the influence of surfactant more quantitatively, in
Fig.~\ref{Fig:sh0fgvar-vs-t} the time dependent lateral domain size $L(t)$
as given in Eq.~\ref{eq:domsize} is shown for a number of surfactant
densities $\rho^s$ between 0.00 and 0.50. Since the lattice is cubic here,
$L(t)$ behaves identically in all three directions.
Fig.~\ref{Fig:sh0fgvar-vs-t}(a) and (b) show identical data, but different
scalings of the axes. In (a), we plot the data linearly in order to give a
better impression of the time dependence of the growth dynamics. However,
in order to check which data is best fitted by the various growth laws, we
provide a log-log scale plot of the same data in
Fig.~\ref{Fig:sh0fgvar-vs-t}(b).  For the first few hundred timesteps, the
randomly distributed fluid densities of the initial system configuration
cause a spontaneous formation of small domains resulting in a steep
increase of $L(t)$. For $\rho^s$=0.00 domain growth does not come to an
end until the domains span the full system. By adding surfactant we can
slow down the growth process and for high surfactant densities $\rho^s >
0.25$, the domain growth stops after a few thousand simulation timesteps.
By adding even more surfactant, the final average domain size becomes very
small and does not grow beyond 7.7 lattice sites.  We fit our numerical
data with the corresponding growth laws and find that for $\rho^s$ smaller
than 0.15 $L(t)$ is best fit by a function proportional to $t^\alpha$. For
$\rho^s$ being 0.15 or 0.20, a logarithmic behaviour proportional to $(\ln
t)^\theta$ is observed. Increasing $\rho^s$ further results in $L(t)$
being best described by a stretched exponential. These results correspond
well with the findings in~\cite{bib:gonzalez-coveney-2}.

\begin{figure}
\centerline{\epsfig{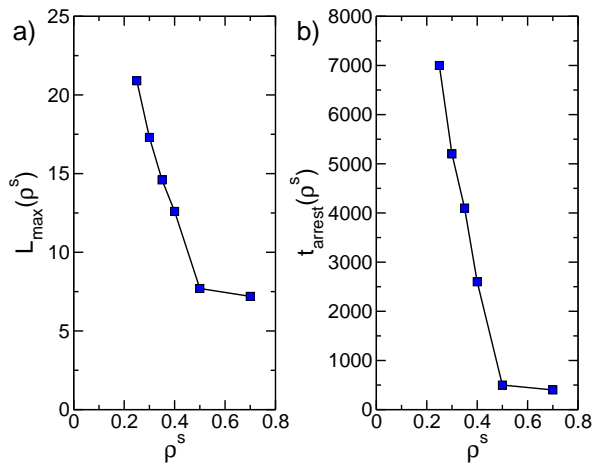}}
%\centerline{\epsfig{file=sh0-dmaxtarrestvsfg.eps,width=0.90\linewidth}}
\caption{\label{Fig:sh0-dmaxtarrestvsfg}(a) Maximum domain size
$L_{\rm max}(\rho^s)$ and (b) time of arrest $t_{\rm arrest}(\rho^s)$ for
various surfactant densities $\rho^s$=0.25, 0.30, 0.35, 0.40, 0.50, 0.7.
$L_{\rm max}(\rho^s)$ as well as $t_{\rm arrest}(\rho^s)$ decrease linearly
with the surfactant density $\rho^s$ and saturate for $\rho^s>0.5$.
All quantities are expressed in lattice units.}
\end{figure}

We study the dependence of the final domain size $L_{\rm max}(\rho^s)$ on
the amount of surfactant as depicted in
Fig.~\ref{Fig:sh0-dmaxtarrestvsfg}(a). It can be observed that the maximum
domain size decreases linearly from 20.9 for $\rho^s = 0.25$ with
increasing $\rho^s$ until a threshold value is reached at $\rho^s=0.5$,
where $L_{\rm max}(t) = 7.7$. Then, $L_{\rm max}(\rho^s)$ decreases much
more slowly and stays almost constant. The slope of the linear regime
corresponds to -52.8. The behaviour of $L_{\rm max}(\rho^s)$ and $t_{\rm
arrest}(\rho^s)$ is consistent with previous lattice
gas~\cite{bib:emerton-coveney-boghosian,bib:love-coveney-boghosian-2001}
and lattice Boltzmann studies~\cite{bib:gonzalez-coveney-2}, where the
authors determine the dependence of the surface tension at a planar
interface between two immiscible fluid species on the surfactant
concentration. In~\cite{bib:gonzalez-coveney-2}, the authors find a linear
dependence between surface tension and surfactant density, but they did
not study such high concentrations as in the current study.
In~\cite{bib:love-coveney-boghosian-2001}, the surface tension approaches
zero for high concentrations, i.e. a saturation occurs causing the size of
the individual fluid domains to saturate as well. These effects can be
explained as follows: adding surfactant to a binary fluid mixture causes
the amphiphiles to minimize the free energy in the system by assembling at
the interface between the two immiscible fluid species. An increase of
surfactant concentration causes the interfacial area to be maximised in
order to accommodate as much surfactant as possible. The increasing
interfacial area causes the individual domains to become smaller and
$L_{\rm max}(\rho^s)$ decreases. If the surfactant concentration becomes
very high ($\rho^s>0.5$ in our case), $L_{\rm max}(\rho^s)$ saturates due
to the maximum possible interfacial area being reached and all available
area being covered with surfactant molecules. More amphiphiles
accumulating at the interface would lead to very steep and energetically
unfavourable gradients of surfactant density in the system. Therefore,
further amphiphiles have to reside within the bulk fluid phases forming micellar
structures. Within our model the minimum final domain size corresponds to
7.7 lattice sites.  However, this value can be varied by tuning the
coupling constants for the amphiphile-amphiphile or amphiphile-fluid
interactions. A more thorough investigation of the influence of the
particular choice of the coupling constants on the final domain size is
of particular interest and shall be investigated within a future project.

In Fig.~\ref{Fig:sh0-dmaxtarrestvsfg}(b), the number of simulation
timesteps $t_{\rm arrest}(\rho^s)$ needed to reach the final domain size
is plotted. Since the time it takes for the system to relax to its
equilibrium state directly depends on the final domain size, it is
consistent with the data presented in Fig.~\ref{Fig:sh0-dmaxtarrestvsfg}(a)
that a linear dependence of $t_{\rm arrest}(\rho^s)$ on the surfactant
concentration can be observed. While for $\rho^s=0.25$ 7000 timesteps are
needed to reach the maximum possible domain size, for $\rho^s=0.5$ 500
timesteps are sufficient. For $\rho^s>0.5$, $t_{\rm arrest}(\rho^s)$
decreases much more slowly than for $\rho^s<0.5$. The slope of $t_{\rm
arrest}(\rho^s)$ in the linear regime is given by -26000. 
%%This result is consistent with Ginzburg-Landau mean field calculations as
%%given in~\cite{bib:gyure-harrington-strilka-stanley:1995} and results
%%obtained from lattice gas simulations~\cite{bib:emerton-coveney-boghosian}.
%- Compare Peter L.: 1/R dependence

\subsection{Steadily sheared systems}
If a binary immiscible fluid mixture is driven mechanically by external
shear forces, it is known that the evolution of domains and phase
separation processes are changed
profoundly~\cite{bib:meyer-2000,bib:zipfel-1999,bib:berghausen-2000}. The
most noticeable effect is the formation of a lamellar phase, i.e. elongated
domains or lamellae form and align along the flow direction. Due to the
anisotropy of the system, the time dependent domain size $L(t)$ behaves
differently for discerned coordinate axes in this case. Furthermore,
modified growth exponents are expected due to the anisotropic effects. 

As already seen in the previous section, adding amphiphiles to a binary
immiscible fluid under shear can change its properties dramatically. The
amphiphiles stabilize the interface between the immiscible fluid species 
and the domain growth is hindered as described in the previous
section. Moreover, the amphiphiles might form complex structures which can
have an impact on the properties of the sheared fluid leading to
non-Newtonian flow~\cite{bib:jens-giupponi-coveney:2006,bib:jens-gonzalez-giupponi-coveney:2005,bib:meyer-2000}.

\begin{figure}
\centerline{\epsfig{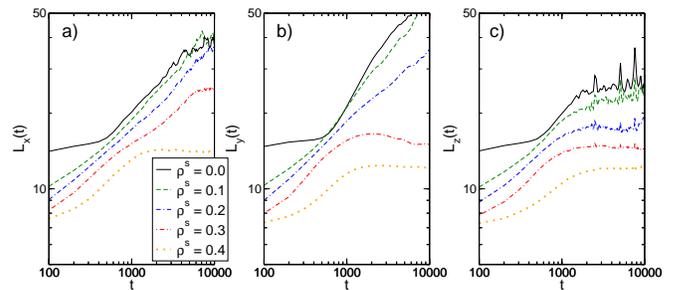}}
%\centerline{\epsfig{file=sh0.1-d_xyz_vs_t-loglog.eps,width=1.00\linewidth}}
%\centerline{\epsfig{file=sh0.1-d_xyz_vs_t.eps,width=0.95\linewidth}}
\caption{\label{Fig:sh0.1-d_xyz_vs_t}(Color online) Domain size
$L(\rho^s)$ in $x$ (a), $y$ (b), and $z$ direction (c) for
surfactant densities $\rho^s$=0.0, 0.1, 0.2, 0.3, 0.4 and a constant shear
rate $\dot\gamma = 1.56\times 10^{-3}$. All quantities are reported in lattice
units.}
% sh=0.1 => 2x0.1/128=0.0015625
\end{figure}

\begin{figure*}
%\centerline{\epsfig{file=Ftotsh01fg02tvar.eps,width=0.9\linewidth}}
\centerline{\epsfig{file=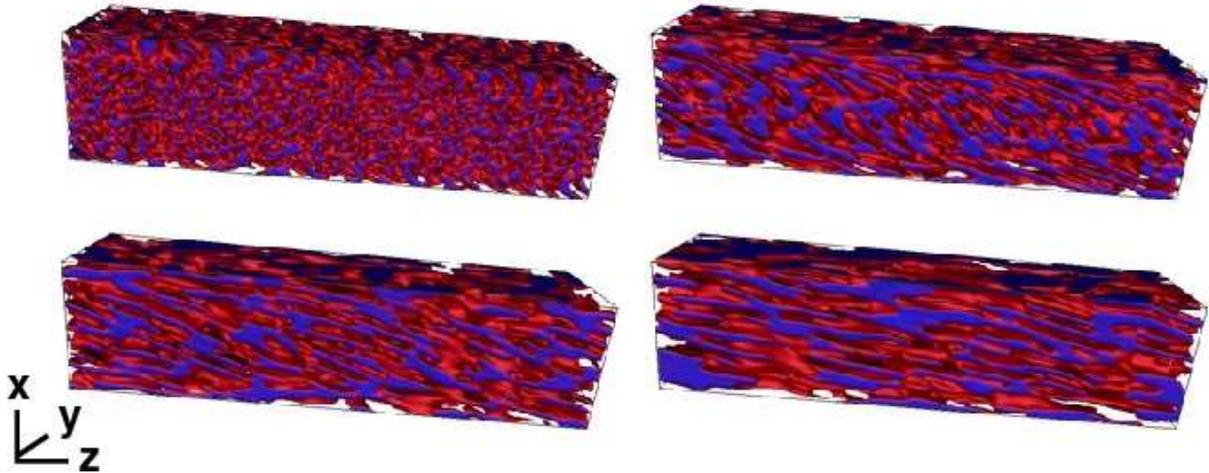,width=0.9\linewidth}}
\caption{\label{Fig:Ftotsh01fg02tvar}(Color online) Volume rendered 
(128x128x512 simulation boxes) fluid densities for surfactant density
$\rho^s$=0.2, a constant shear rate $\dot\gamma = 1.56\times 10^{-3}$ and
variable number of time steps $t$=1000 (upper left), $t$=4000 (upper
right), $t$=6000 (lower left), and $t$=10000 (lower right). While the
shape of individual domains does not show distinct features at early times
of the simulation, elongated structures appear at $t$=4000 and start to
become aligned with the shear at $t$=6000. At late stages of the
simulation run ($t$=10000), the lamellar phase characterised by thin and
long lamellae filling the whole system can be observed.
}
\end{figure*}

We study ternary 128x128x512 sized systems under constant
shear. The shear rate is set to $\dot\gamma =1.56\times10^{-3}$ and
$\dot\gamma = 3.12\times 10^{-3}$, while the surfactant density is
varied between $\rho^s$=0.0 and 0.4. 
Fig.~\ref{Fig:sh0.1-d_xyz_vs_t} shows the time dependent lateral domain
size for all three coordinate axes at $\dot\gamma = 1.56\times 10^{-3}$.
In the $x$ direction, which is the axis between the shear planes, the power
law regime of $L_x(t)$ starts at $t$=500 for the $\rho^s$=0.0 case, while
for higher $\rho^s$ the initial growth regime is overcome by the power law
regime before the first measurement at $t$=100. As long as $\rho^s < $
0.3, the growth rate is not hindered by the amphiphiles and domains grow until the end of the simulation. For
$\rho^s$=0.3 the power law regime starts to break down at $t$=900 and
$L_x(t)$ saturates at $t$=5000. Adding even more surfactant results in an
even earlier saturation at $t$=1500. The $y$ direction is the direction
parallel to the shear planes and perpendicular to the direction of shear.
Since this direction is less affected by the shear forces, $L_y(t)$ grows
faster than $L_x(t)$ for low surfactant concentrations
$\rho^s<$0.2 causing the domains forming being extended in the $y$ direction.
For $\rho^s$=0.2 $L_x(t)$ and $L_y(t)$ behave almost identical, while for
$\rho^s>$0.2 a crossover occurs and the maximum attainable value for
$L_y(t)$ is below the result for $L_x(t)$. In the direction of shear ($z$
direction), $L_z(t)$ saturates even for the no surfactant case at
$L_z(t)$=25 and comes to arrest at even smaller values with increasing $\rho^s$.
The complex behaviour of $L_i(t)$ can be better understood by reminding
ourselves that the domain size is measured in the direction of the
Cartesian coordinate axes. However, individual fluid domains occurring in
the system are being elongated due to the shear and try to align with the
shear gradient. Thus, they are not parallel to any coordinate axis.
Therefore, with a measurement of $L_z(t)$ one is not able to detect the
actual length of individual lamellae, but only their thickness in the $z$
direction. Similar arguments are valid for $L_x(t)$, shear causing the
measured domain size in the $x$ direction to be larger than the lamellae's
thickness.  For increasing $\rho^s$, the average domain size reduces due
to the influence of the amphiphiles, thus causing the individual domains
to become smaller. If $\rho^s>$0.2, the alignment of the domains with the
shear causes $L_x(t)$ to appear larger than $L_z(t)$. For high surfactant
concentrations ($\rho^s$=0.4) all three directions behave very similarly:
domain growth comes to an end after less than 2000 timesteps and the final
domain size is between 10 and 15 lattice units in all directions
, signaling the appearance of a stable microemulsion where the shape of the
domains is almost unaffected by the shear.

Regular peaks occur in $L_z(t)$ at every 2500 timesteps with less
pronounced peaks in between them. These peaks can be explained as follows:
For the stretching of domains, a certain amount of work against surface
tension is needed. On macroscopic scales, the stress tensor does not
vanish due to the viscoelastic response of the
system~\cite{bib:rothman:1991,bib:krall-sengers-hamano:1992}. On the
microscale, however, a breakup and recombination of domains can be
observed~\cite{bib:ohta-nozaki-doi:1990}. These domains grow by diffusion
and eventually join each other to form larger structures. If the internal
stress becomes too large due to the shear induced deformation, they break
up and start to form again. Assuming a large system with many independent
domains growing and breaking incoherently, the only observable effect
might be a slowing down of the domain growth. In contrast, if the growth
and breakup occur coherently as they do in our simulations, a periodicity
in the measured time dependent domain size can be
observed~\cite{bib:corberi-gonnella-lamura}. As can be observed in  
Fig.~\ref{Fig:sh0.1-d_xyz_vs_t}c), the frequency of domain breakup is
independent of the surfactant concentration, while the height of the peaks
decreases with increasing $\rho^s$.

%FIGVOLREND: The peaks occur due to individual lamellae joining each
%other in order to form a larger one. However, if these lamellae are
%crossing the periodic boundaries, they cannot be stable. Instead, they
%break up immediately and start to form again. In our simulations these
%finite size effects appear on a scale of about 2500 timesteps causing a
%strong peak in $L_z(t)$. Such effects have already been reported for
%lattice gas simulations~\cite{bib:love-coveney:2002}.
%
%For domain sizes which are greater than one fourth of the system size, we
%cannot be sure that our simulations are free of finite size effects
%anymore, i.e. it is not possible to state with certainty if the deviation
%of $L_x(t)$ for $\rho^s=0.0$ and $0.1$ and $t>$4000 is a correct
%observation or an artefact\fixme{put this somewhere else}

Figure~\ref{Fig:Ftotsh01fg02tvar} shows volume rendered examples of a
simulated system with surfactant density $\rho^s$=0.2 and a constant shear
rate of $\dot\gamma = 1.56\times 10^{-3}$. The four snapshots are taken a
different times $t$=1000 (upper left), $t$=4000 (upper right), $t$=6000
(lower left), and $t$=10000 (lower right). It can be observed that at
early stages of the simulation, the shape of individual domains does not
show distinct features, while at $t$=4000, slightly elongated domains
start to occur which begin to align with the shear gradient. At $t$=6000,
these features are substantially more dominant and at late simulation
times ($t$=10000) the system is filled with elongated and thin lamellae
consisting of one of the immiscible fluid species and which are almost
parallel to the shear plane. 

\begin{figure}
\centerline{$\mathbf{\rho^s=0.0}$\qquad\qquad\qquad\qquad\qquad\qquad$\mathbf{\rho^s=0.3}$\vspace{-0.4cm}}
%\centerline{a)\vspace*{-0.4cm}\epsfig{file=colour_Ftot1.6_128cub_sh0.1fg0_t010000.sf3dxsum.pre.eps,width=0.40\linewidth,angle=270}\hspace{-0.1\linewidth}d)\epsfig{file=colour_Ftot1.6_128cub_sh0.1fg0.3_t010000.sf3dxsum.pre.eps,width=0.40\linewidth,angle=270}}
%\centerline{b)\vspace*{-0.4cm}\epsfig{file=colour_Ftot1.6_128cub_sh0.1fg0_t010000.sf3dysum.pre.eps,width=0.40\linewidth,angle=270}\hspace{-0.1\linewidth}e)\epsfig{file=colour_Ftot1.6_128cub_sh0.1fg0.3_t010000.sf3dysum.pre.eps,width=0.40\linewidth,angle=270}}
%\centerline{c)\epsfig{file=colour_Ftot1.6_128cub_sh0.1fg0_t010000.sf3dzsum.pre.eps,width=0.40\linewidth,angle=270}\hspace{-0.1\linewidth}f)\epsfig{file=colour_Ftot1.6_128cub_sh0.1fg0.3_t010000.sf3dzsum.pre.eps,width=0.40\linewidth,angle=270}}
%
\centerline{a)\vspace*{-0.4cm}\epsfig{file=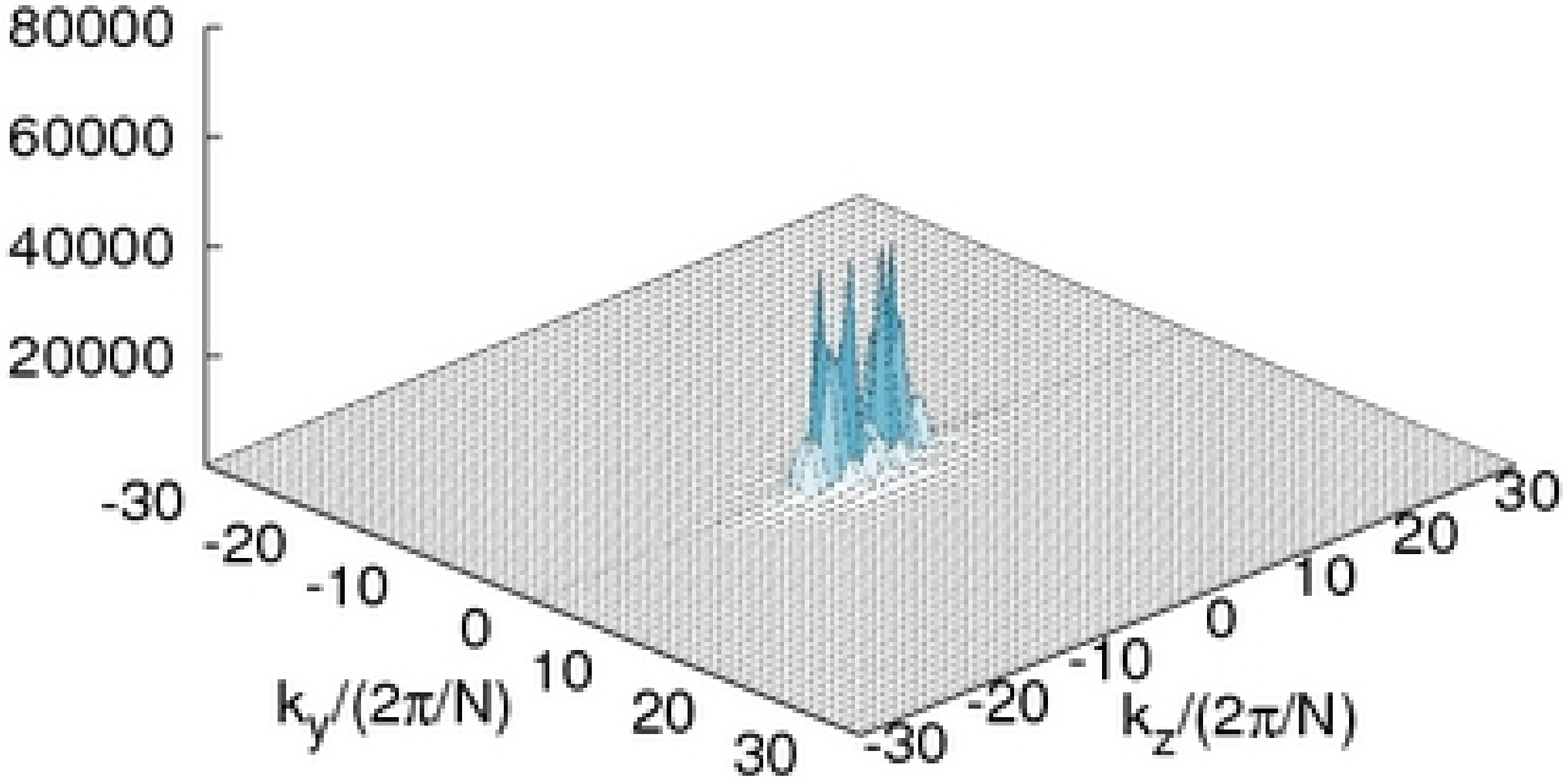,width=0.40\linewidth,angle=270}\hspace{-0.1\linewidth}d)\epsfig{file=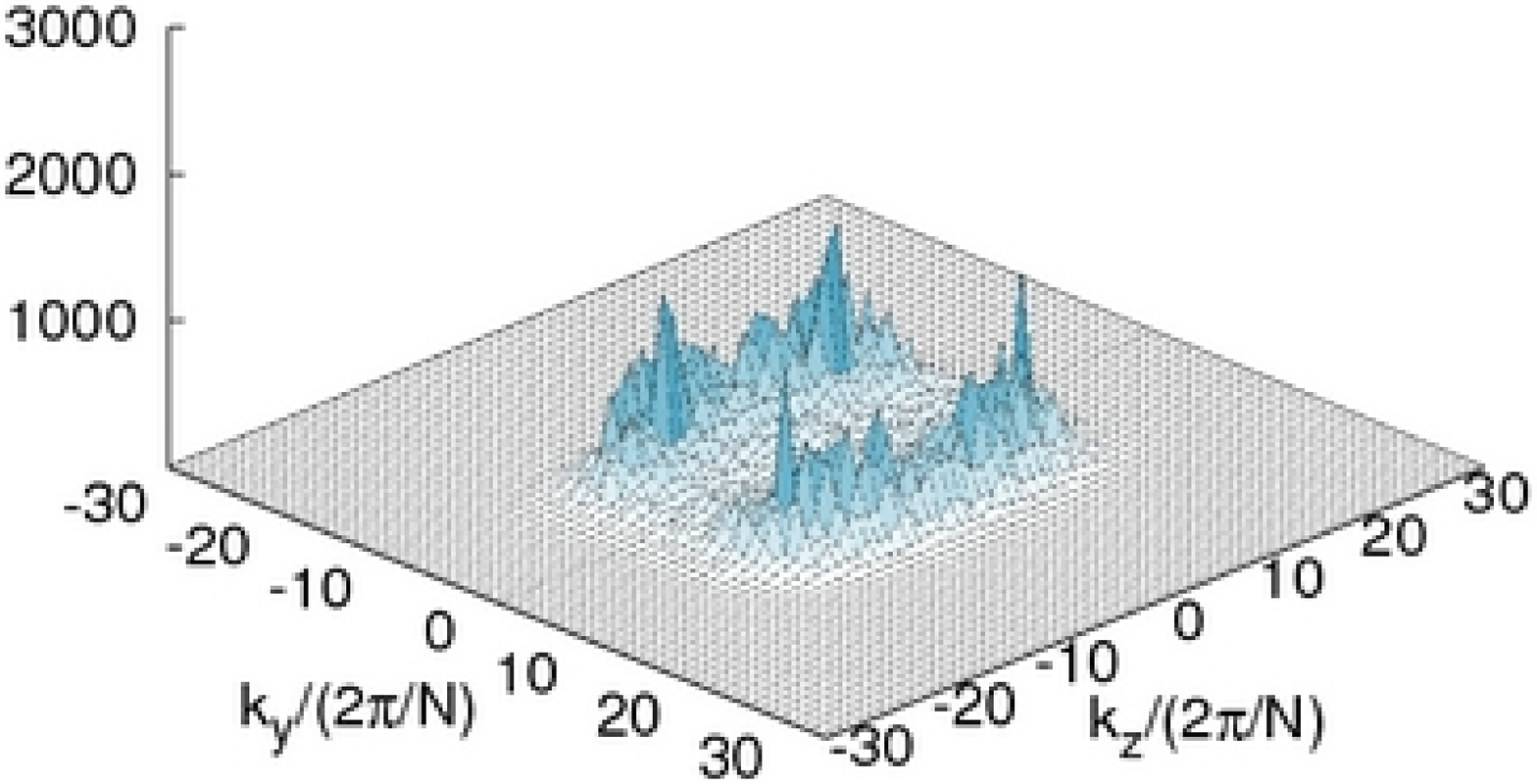,width=0.40\linewidth,angle=270}}
\centerline{b)\vspace*{-0.4cm}\epsfig{file=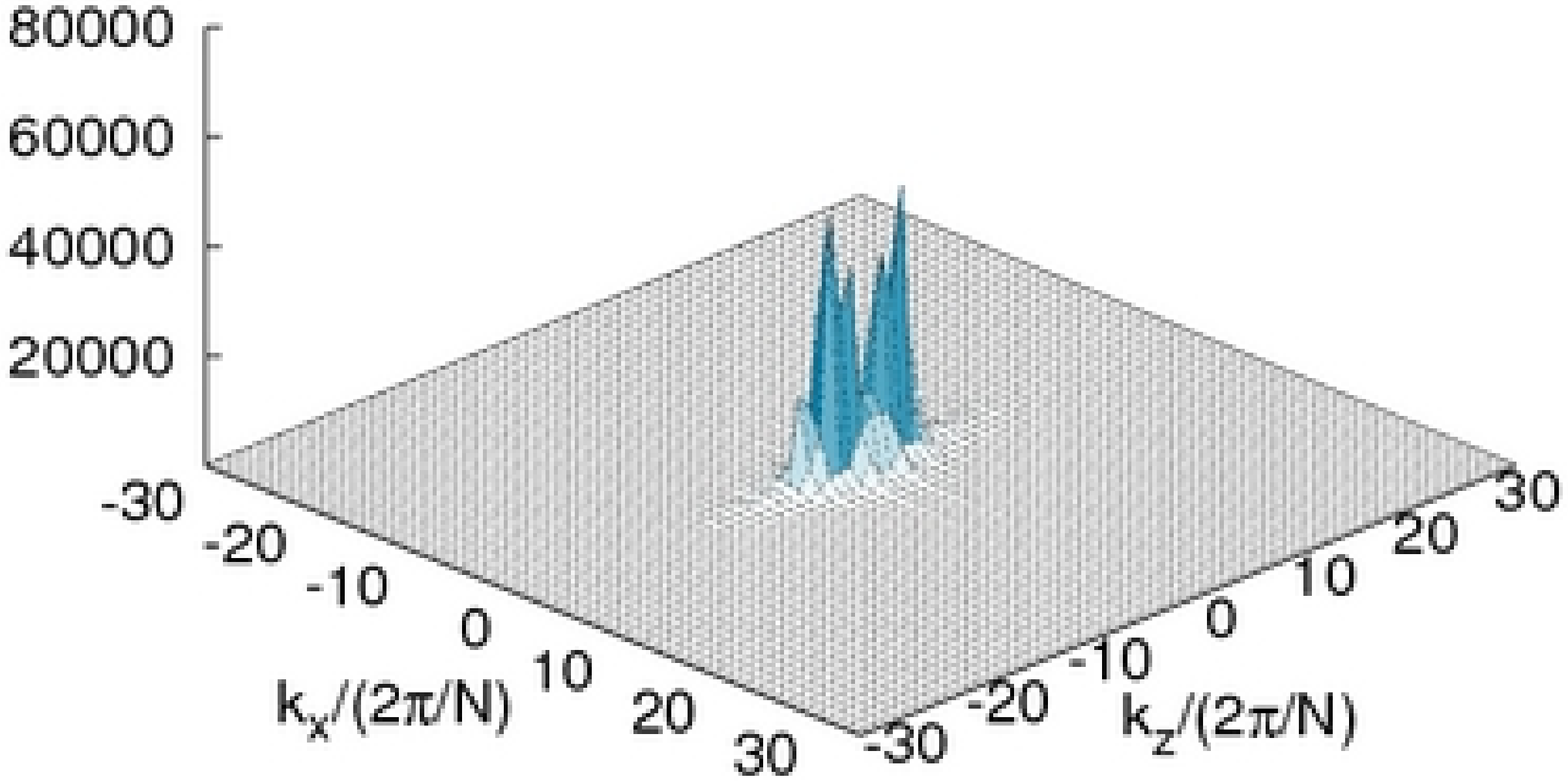,width=0.40\linewidth,angle=270}\hspace{-0.1\linewidth}e)\epsfig{file=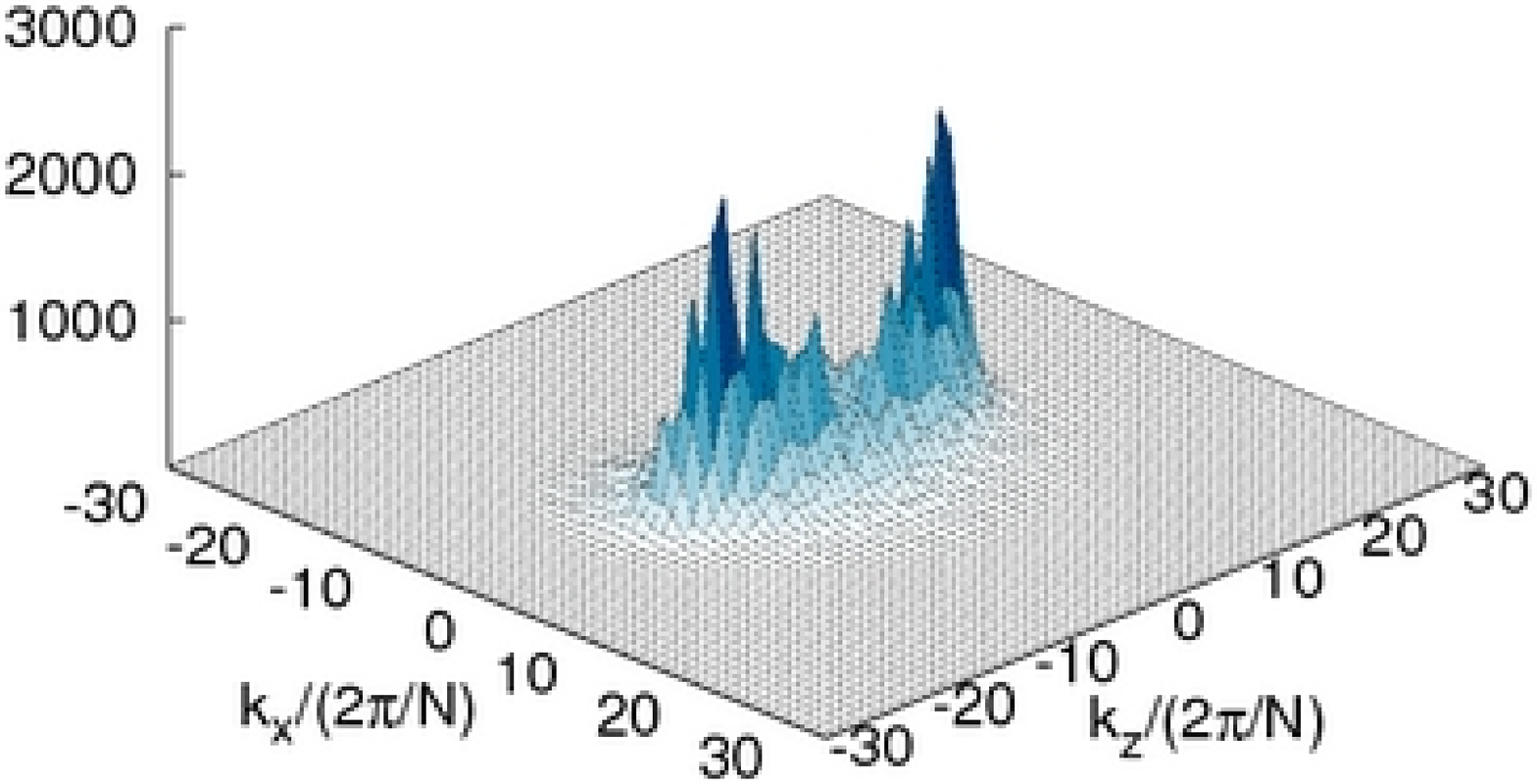,width=0.40\linewidth,angle=270}}
\centerline{c)\epsfig{file=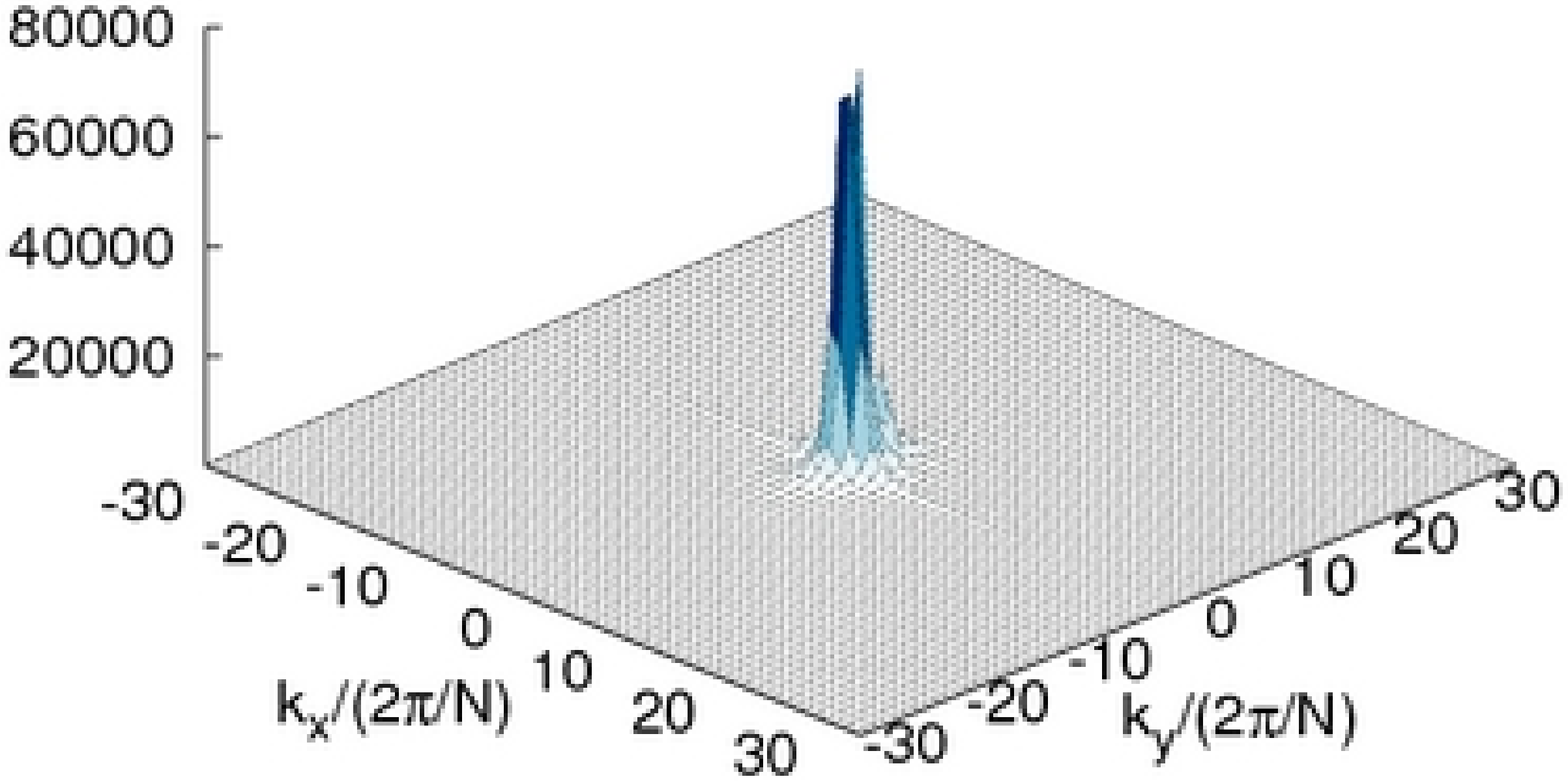,width=0.40\linewidth,angle=270}\hspace{-0.1\linewidth}f)\epsfig{file=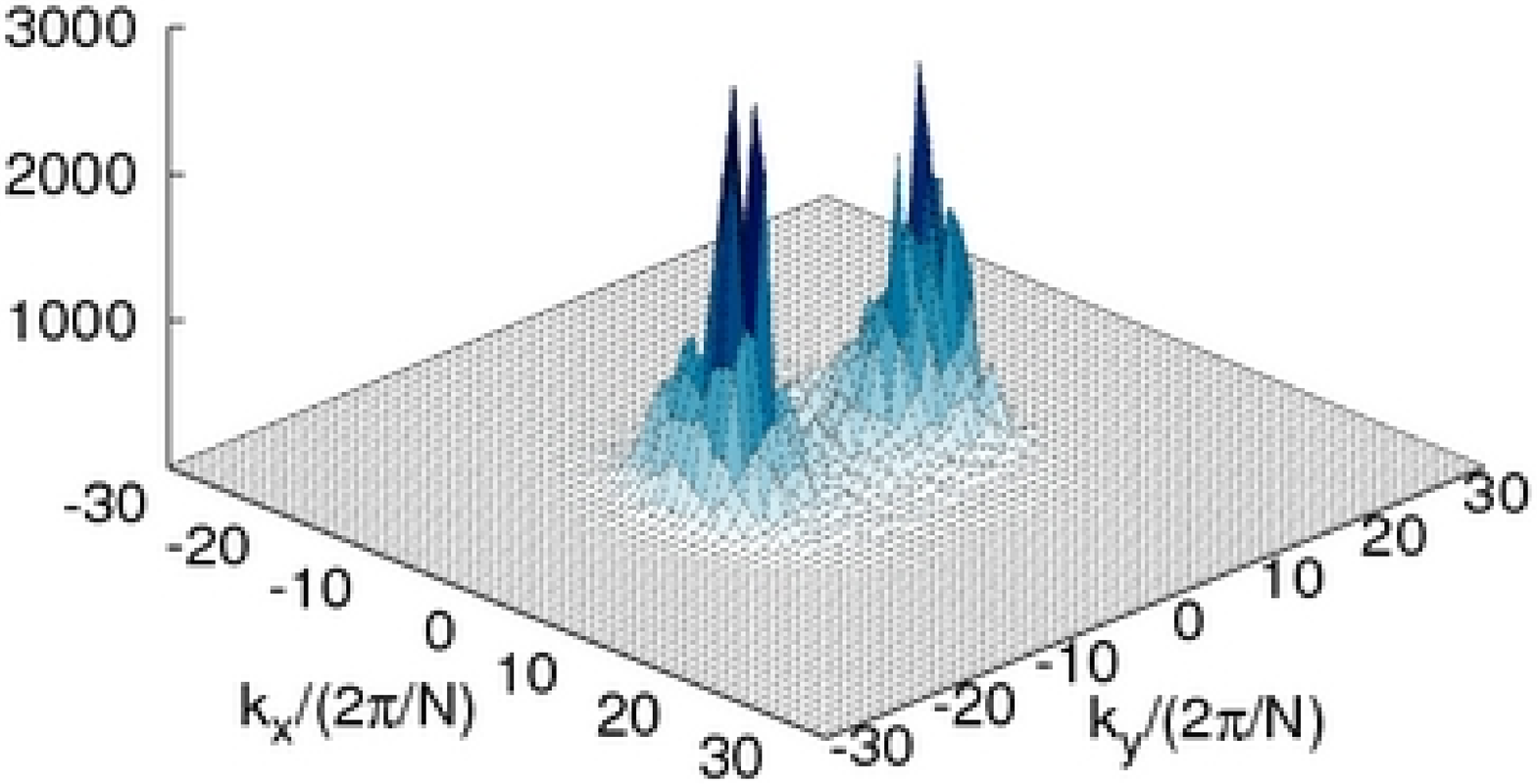,width=0.40\linewidth,angle=270}}
\caption{\label{Fig:SFsh0.1fgvar}(Color online) Projected structure functions
(``scattering pattern'') $S_i$ for surfactant concentrations $\rho^s=0.0$
(a-c) and $\rho^s=0.3$ (d-f) at $t=10000$. Projections are in $x$-, $y$-
and $z$ directions (from top to bottom) and for a cubic cutout
of the full system with side length $N$=128.}
\end{figure}

In order to permit a comparison with experimentally available scattering
data, in Fig.~\ref{Fig:SFsh0.1fgvar} we present projected structure
functions for the surfactant-less case (a-c) and a surfactant density of
$\rho^s=0.30$ (d-f) at $t=10000$. The $x$, $y$ and $z$ directions are
shown from top to bottom. The shown projections are for a cubic 128$^3$
cutout of the elongated systems. In contrast to the non-sheared case, all
three directions show distinguished properties: For $\rho^s=0.00$, a high
peak of $S_z(k_x,k_y,t)$ can be observed around $k_x=k_y=0$, while in $x$-
and $y$ direction two less high peaks at positions $>0$ show up. These
data can be interpreted as follows: At $t=10000$, the domain size in the
direction of the flow corresponds to the size of the cubic cutout, i.e.
128 lattice sites. In the $x$ and $y$ directions, however, the size of
the occurring structures is smaller, indicating the occurrence of very long
lamellar structures in the system.  By adding surfactant to the system
(Fig.~\ref{Fig:SFsh0.1fgvar}(d-f)), the occurring length scales depicted
decrease by the splitting of the single peak in the $z$-projection and the
occurrence of two small peaks at $k_x=0$ and $k_y=\pm10$. While the peaks
in the $y$ direction denote similar length scales as in the $z$ direction, the
projections in the direction between the shear planes ($x$) show a
different behaviour. Here, two parallel structures at $k_y=\pm 10$ and
$k_z$ between -20 and 20 indicate a much wider variation of the thickness
of the individual domains. This is in contrast to the non-sheared case in
Fig.~\ref{Fig:SFsh0fgvar}, where a volcano-like shape of the structure
factor was observed. 

\begin{figure}
%\centerline{\epsfig{file=sh0.2-d_xyz_vs_t-loglog.eps,width=1.0\linewidth}}
\centerline{\epsfig{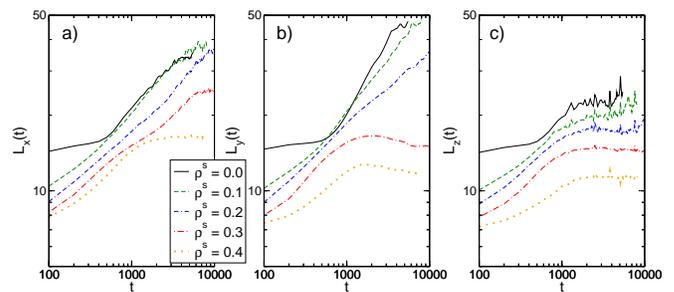}}
\caption{\label{Fig:sh0.2-d_xyz_vs_t}(Color online) Domain size
$L(\rho^s)$ in (a) $x$-, (b) $y$-, and (c) $z$ direction for
surfactant densities $\rho^s$=, 0.1, 0.2, 0.3, 0.4 and a constant shear
rate $\dot\gamma = 3.12\times 10^{-3}$.}
% sh=0.2 => 2x0.2/128=0.003125
\end{figure}

Doubling the shear rate to $\dot\gamma = 3.12\times 10^{-3}$ results in
very similar behaviour, as shown in Fig.~\ref{Fig:sh0.2-d_xyz_vs_t}.
%However, in the $x$ direction it can be observed that $L_x(t)$ behaves
%almost identically for $\rho^s=0.0$ and $0.1$ at $t>500$. This effect
%suggests that shear is able to limit the effect of surfactant and/or to
%counteract the decrease of surface tension due to the present amphiphiles. 
In the $z$ direction, peaks can now be observed even for $\rho^s=0.4$, but
$L_z(t)$ is much more noisy for lower surfactant concentrations. However,
it can still be seen that there is a number of equidistant peaks for
$\rho^s<0.4$ which occur every 2500 timesteps with some additional peaks
in between in the case of $\rho^s=0.0$ and $0.1$. The equidistant peaks
occurring with the same frequency as in the $\dot\gamma = 1.56\times
10^{-3}$ case shows that the breakup phenomena observed are independent of
the shear rate.

A number of experiments have reported that the shear stabilizes the system
and causes the phase separation to come to an end with $L_z(t)$ being very
large and $L_x(t)$ being much
smaller~\cite{bib:hashimoto-matsuzaka-moses-onuki-1994,bib:shou-chakrabarti:2000,bib:stansell-stratford-desplat-adhikari-cates:2006,bib:wagner-yeomans}.
We are not able to reproduce these results due to the limited size of our
simulations: substantially larger systems and higher shear rates need to
be studied in order to quantify the arrest of domain growth due to shear.
However, the computational resources needed would be at the limit of what
can be done on current supercomputers.
   
We have shown in this section that the dynamical scaling hypothesis does
not hold for sheared ternary systems in three dimensions since we indeed
find three individual length scales pointing out the transition from the
sponge to the lamellar phase: while in the flow direction ($z$), $L(t)$ is
determined by the resultant length of the occurring lamellae, in the
direction between the shear planes ($x$), the domains grow steadily and
exhibit power law behaviour up to a maximum that depends on the surfactant
concentration. In the $y$ direction, domain growth is not hindered by
shear. In fact, $L_y(t)$ grows slightly faster than in the non-sheared
case. Increasing the surfactant concentration has a strong impact on
domain growth: starting at $\rho^s$=0.3, $L_y(t)$ and $L_z(t)$ recover the
behaviour of the case without shear, i.e. the length scales saturate
around 15. In the $x$ direction, however, growth continues to up to
$L_x(t)$=26. This can be explained as follows: with increasing surfactant
concentration, the final domain sizes become smaller, reducing the
influence of the shear forces in the $y$ and $z$ directions. In the
direction between the shear planes, however, an increase of $L_x(t)$ can
still be observed because the domains are still being elongated due to
shear and try to align with the velocity profile. Thus, they are tilted
and their size appears to be smaller than it actually is in $z$ direction
and larger in the $x$ direction.

\begin{figure*}
%\centerline{\epsfig{file=Ftotsh01w0001fg02tvar.eps,width=0.9\linewidth}}
\centerline{\epsfig{file=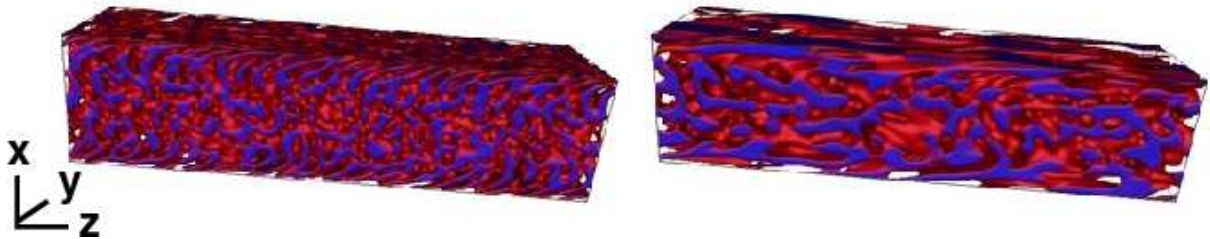,width=0.9\linewidth}}
\caption{\label{Fig:Ftotsh01w0001fg02tvar}(Color online) Volume rendered fluid
densities for surfactant density $\rho^s$=0.2 at $t$=2500 (left) and
$t$=10000 (right). The shear rate is $\dot\gamma = 1.56\times 10^{-3}$ and
$\omega = 0.001$. At $t$=2500, the shear velocity is close to its reversal
point and the domains are aligned vertically close to
the shear plane, while in the bulk of the system no preferred orientation
can be observed. After 10000 timesteps, however, the domains close to the
shear plane are aligned with the flow direction because the shear planes
are in a position just before their reversal point. In the bulk, still no
preferred orientation can be found since it takes longer for the
velocity gradient to penetrate the whole system than the duration of a
single period of shear.}
\end{figure*}

Our findings are in agreement with Ginzburg-Landau and Langevin
calculations~\cite{bib:corberi-gonnella-lamura:1998,bib:corberi-gonnella-lamura:1999,bib:corberi-gonnella-lamura,bib:corberi-gonnella-lamura:2002}
as well as two-dimensional lattice-Boltzmann simulations of binary
immiscible fluid mixtures as presented
in~\cite{bib:gonnella-orlandini-yeomans,bib:xu-gonnella-lamura:2004,bib:stansell-stratford-desplat-adhikari-cates:2006}.
However, to the best of our knowledge, there are no detailed theoretical
studies of the dependence of domain growth properties on the surfactant
concentration.  The only known work utilizes a Ginzburg-Landau free-energy
approach to study sheared microemulsions, but does not vary the amount of
surfactant.  In addition, the authors only cover two-dimensional systems
and are thus unable to describe the behaviour of
$L_y(t)$~\cite{bib:corberi-gonnella-suppa:2001}.

\subsection{Complex fluids under oscillatory shear}
In the case of oscillatory shear, the morphology and the domain growth are
altered significantly, although the average deformation is zero after each
period of shear. For example, it has been found experimentally for binary
fluid mixtures that for very low oscillation frequencies domain growth can
be interrupted~\cite{bib:krall-sengers-hamano:1993}, or domains can grow
on much longer time scales than given by the oscillation
frequency~\cite{bib:matsuzaka-jinnai-etal:1997}. Simulations so far either
do not include hydrodynamic effects, or are limited to two
dimensions~\cite{bib:xu-gonnella-lamura:2003}. It has been observed in the
two-dimensional lattice Boltzmann studies by Xu et al. that hydrodynamic
effects must not be neglected in the case of oscillatory
shear since there exists a finite time inversely proportional to the
viscosity which is required to set a linear velocity profile in the
system. For high oscillation
frequencies, this time is longer than the oscillation period and there
will never be a linear velocity profile in the system, thus influencing
the domain growth substantially~\cite{bib:xu-gonnella-lamura:2003}.

Due to its higher complexity, oscillatory shear has much less been a
subject of research projects than systems under steady shear. Therefore,
the number of publications that can be found in the literature is
substantially smaller. To the best of our knowledge, no detailed
three-dimensional simulation studies of phase separation in ternary
amphiphilic fluid mixtures under oscillatory shear have been reported on
so far. The most detailed three-dimensional simulations we are aware of
are our own studies of the gyroid mesophase under oscillatory shear as
presented in~\cite{bib:jens-giupponi-coveney:2006}.

In this section we present our results of simulations of systems
equivalent to the ones considered in the previous section, but with the
shear plane moved as given by Eq.~\ref{coseno}. We apply two
different oscillation frequencies $\omega = 0.001$ and $\omega = 0.01$,
where a single oscillation takes 6283 timesteps in the slow case and 628.3
timesteps in the fast case. 

\begin{figure}
%\centerline{\epsfig{file=sh0.1w0.001-d_xyz_vs_t-loglog.eps,width=1.0\linewidth}}
\centerline{\epsfig{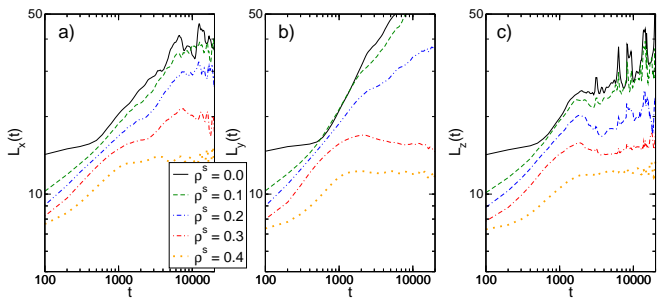}}
\caption{\label{Fig:sh0.1w0.001-d_xyz_vs_t}(Color online) Domain size
$L(\rho^s)$ in $x$- (a), $y$- (b), and $z$ direction (c) for
surfactant densities $\rho^s$=, 0.1, 0.2, 0.3, 0.4 and oscillatory shear
with $\dot\gamma = 1.56\times 10^{-3}$, $\omega = 0.001$.}
\end{figure}
Let us first consider the case with a lower oscillation frequency and
lower shear rate, i.e.
$\omega = 0.001$ and $\dot\gamma = 1.56\times 10^{-3}$. In the case of oscillatory sheared systems, the
individual fluid domains try to align with the velocity gradient as in the
previous section. However, since we do not consider steadily moving shear
planes here, domains are never able to reach a steady state and
instead have to follow the oscillation of the planes. The frequencies
considered in our simulations are comparably high since no linear velocity
gradient sustains long enough during a single oscillation for the domains
to fully align with it. This is depicted in
Fig.~\ref{Fig:Ftotsh01w0001fg02tvar} which shows two typical examples from
a simulation with $\rho^s$=0.2. On the left hand side, a volume rendered
snapshot is given at $t$=2500. Here, the oscillating shear planes have
just passed their reversal point. Close to the shear planes, the domains
are aligned vertically because they have to be turned around in order to
follow the changing direction of movement of the shear planes. In the
bulk of the system, however, no preferred direction can be observed since
the velocity gradient does not interpenetrate the whole system. At
$t$=10000, the shear planes are in a position just before their reversal
point. Thus, the fluid mixture was accelerated for more than 2000
timesteps and the domains close to the shear boundary are well aligned in
the direction of the flow. In the bulk, again no preferred direction can
be observed.

The time dependent lateral domain size of this simulation and for varied
surfactant concentration is presented in
Fig.~\ref{Fig:sh0.1w0.001-d_xyz_vs_t}. As in the case of continuous shear,
the domain growth in $y$ direction is almost uninfluenced by the applied
shear forces. In fact, for low surfactant concentrations $\rho^s < 0.2$,
$L_y(t)$ even grows slightly faster than in the case without oscillatory
movement. For $\rho^s\ge 0.2$ the maximum domain size obtained is similar
to the steady shear case. Due to the non-steady movement of the shear
planes, $L_x(t)$ and $L_z(t)$ show a more rich behaviour: both functions
show distinct kinks around the reversal points of the shear and for
$\rho^s\ge 0.2$ it is found that the growth rates are smaller than in the
case of steady shear. Thus, we can observe the formation of tubular
structures which are elongated in the $y$ direction and show similar length
scales in $x$ and $z$ direction. For very high surfactant concentrations
($\rho^s$=0.4) it is not possible to distinguish between tubular and
spherical structures due to the small size of the individual domains.

In the $z$ direction we can still observe peaks related to the formation
and breakup, as well as the rotational movement, of domains. However, due
to the overlaid effect of the oscillation, these peaks are not
equidistant as in the steady shear case anymore.

\begin{figure}
%\centerline{\epsfig{file=sh0.1w0.01-d_xyz_vs_t-loglog.eps,width=1.0\linewidth}}
\centerline{\epsfig{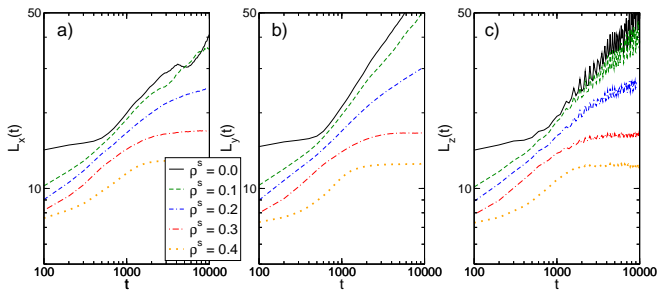}}
\caption{\label{Fig:sh0.1w0.01-d_xyz_vs_t}(Color online) Domain size
$L(\rho^s)$ in $x$ (a), $y$ (b), and $z$ direction (c) for
surfactant densities $\rho^s$=, 0.1, 0.2, 0.3, 0.4 and oscillatory shear
with $\dot\gamma = 1.56\times 10^{-3}$, $\omega = 0.01$.}
\end{figure}

In Fig.~\ref{Fig:sh0.1w0.01-d_xyz_vs_t} we present a system with a ten
times larger oscillation frequency, i.e. $\omega = 0.01$.  Here, the
frequency of the oscillations is so high that the fluid is not able to
follow the movement of the walls anymore. Thus, the influence of the shear
on the growth behaviour becomes less pronounced, with the domains
constantly growing as long as the amount of surfactant present allows it.
The growth rates are comparable to the non-sheared case here and show
identical growth laws as in the non-sheared case. The only difference is
that the exponents are found to be smaller while  $L_y(t)$ grows
slightly faster than $L_x(t)$ and $L_z(t)$ depicting the occurrence of
tubular structures in the system. The $z$ direction is the only
component of the time dependent lateral domain size that differs from the
unsheared case, because strong oscillations start to appear due to the
distortions caused by the moving boundaries.

We have increased the shear rate to $\dot\gamma = 3.12\times 10^{-3}$, but
the lateral domain sizes obtained are almost identical to the $\dot\gamma
= 1.56\times 10^{-3}$ case. Therefore, we do not present an additional
figure, but our findings can be used to argue that for high oscillation
frequencies, the system behaves in a manner equivalent to the non-sheared case. 

In the case of oscillatory shear we have shown the occurrence of tubular
structures due to shear imposed anisotropic domain growth, the slowing
down of the domain growth rate depending on the oscillation frequency, as
well as that a microemulsion with high surfactant concentration stays
unaffected by external shear forces. Our results are in agreement with the
simulations of Qiu et al.~\cite{bib:qiu-zhang-yang:1998} and the two
dimensional lattice Boltzmann simulations of Xu et
al.~\cite{bib:xu-gonnella-lamura:2003}.

%genauer, mehr refs, besser erklären was wir daraus lernen: weniger
%deskriptiv.
\section{Conclusions}
In this paper we have presented for the first time detailed three-dimensional
lattice Boltzmann studies of binary immiscible and ternary amphiphilic fluid
mixtures under constant and oscillatory shear. 

We have reproduced the well-known power law growth of domains in the case of
binary immiscible fluids (spinodal decomposition) which crosses over to a
logarithmic law and to a stretched exponential if one increases the surfactant
concentration even further. For sufficiently high surfactant concentrations,
domain growth can come to an end and the system corresponds to a stable
bicontinuous microemulsion. For amphiphile concentrations of up to 30\% we find
linear dependencies of the time of arrest as well as the maximum domain size on
the amphiphile concentration. For concentrations above 30\%, both arrest length
and time of arrest do not change any more since the surface tension at the
interfaces between the two immiscible fluids is at its minimum.  The whole
interface is filled with amphiphiles and further amphiphile molecules have to
reside within the bulk fluid. 

In sheared systems, we have studied the influence of moving boundaries on the
effect of domain growth and report domain breakup phenomena depending on the
shear rate as well as the amphiphile concentration. Depending on the surfactant
concentration and the shear rate, we find a transition from a sponge phase to a
lamellar phase.

Under oscillatory shear and with the oscillations frequencies chosen, no linear
velocity gradient can build up within a single period of shear. Thus, the
domains are constantly rearranging and align with the flow in the vicinity of
the shear planes. In the bulk, however, no preferred alignment can be observed.
But since the growth in the $y$ direction is not hindered by the shear, tubular
structures occur and are best observable for low surfactant concentrations. For
very fast oscillations ($\omega = 0.01$), the system is not able to
follow the external shear at all. Thus, it behaves similar to a non-sheared one. The
only differences are that fluctuations in $L_z(t)$ can be observed due to the
oscillatory forces, while the growth in the other two directions is slowed
down. For surfactant concentrations $\rho^s \le 0.2$ anisotropic growth in $x$-
and $y$ direction is observed depicting the presence of tubular domains in the
system. In future work it would be of interest to study the formation of tubes
and their dependency on the shear rate, oscillation frequency and surfactant
concentration in greater detail. Additionally, the study of asymmetric
mixtures, where the concentrations of the two immiscible fluid species differs,
remains unexplored. 

%\ack
\begin{acknowledgments}
We are grateful for the support of the HPC-Europa programme, funded under the
European Commission's Research Infrastructures activity, contract number
RII3-CT-2003-506079. We also acknowledge RealityGrid and ESLEA EPSRC grants
(GR/R67699 and GR/T04465), the H\"ochstleistungsrechenzentrum Stuttgart for
providing access to their NEC SX8 and the National Science Foundation (NSF) for
US TeraGrid usage within NRAC grant MCA04N014.  We would like to thank
H.~Berger, R.~Keller, and P.~Lammers for technical support and K.~Stratford for
fruitful discussions.
\end{acknowledgments}
%\bibliography{main,jens-pub}

\begin{thebibliography}{72}
\expandafter\ifx\csname natexlab\endcsname\relax\def\natexlab#1{#1}\fi
\expandafter\ifx\csname bibnamefont\endcsname\relax
  \def\bibnamefont#1{#1}\fi
\expandafter\ifx\csname bibfnamefont\endcsname\relax
  \def\bibfnamefont#1{#1}\fi
\expandafter\ifx\csname citenamefont\endcsname\relax
  \def\citenamefont#1{#1}\fi
\expandafter\ifx\csname url\endcsname\relax
  \def\url#1{\texttt{#1}}\fi
\expandafter\ifx\csname urlprefix\endcsname\relax\def\urlprefix{URL }\fi
\providecommand{\bibinfo}[2]{#2}
\providecommand{\eprint}[2][]{\url{#2}}

\bibitem[{\citenamefont{Bray}(1994)}]{bib:bray}
\bibinfo{author}{\bibfnamefont{A.~J.} \bibnamefont{Bray}},
  \bibinfo{journal}{Adv. Phys.} \textbf{\bibinfo{volume}{43}},
  \bibinfo{pages}{357} (\bibinfo{year}{1994}).

\bibitem[{\citenamefont{Gonz\'{a}lez-Segredo
  et~al.}(2003)\citenamefont{Gonz\'{a}lez-Segredo, Nekovee, and
  Coveney}}]{bib:gonzalez-nekovee-coveney}
\bibinfo{author}{\bibfnamefont{N.}~\bibnamefont{Gonz\'{a}lez-Segredo}},
  \bibinfo{author}{\bibfnamefont{M.}~\bibnamefont{Nekovee}}, \bibnamefont{and}
  \bibinfo{author}{\bibfnamefont{P.~V.} \bibnamefont{Coveney}},
  \bibinfo{journal}{Phys. Rev. E} \textbf{\bibinfo{volume}{67}},
  \bibinfo{pages}{046304} (\bibinfo{year}{2003}).

\bibitem[{\citenamefont{Kawakatsu et~al.}(1993)\citenamefont{Kawakatsu,
  Kawasaki, Furusaka, Obayashi, and Kanaya}}]{bib:kawakatsu-kawasaki-1993}
\bibinfo{author}{\bibfnamefont{T.}~\bibnamefont{Kawakatsu}},
  \bibinfo{author}{\bibfnamefont{K.}~\bibnamefont{Kawasaki}},
  \bibinfo{author}{\bibfnamefont{M.}~\bibnamefont{Furusaka}},
  \bibinfo{author}{\bibfnamefont{H.}~\bibnamefont{Obayashi}}, \bibnamefont{and}
  \bibinfo{author}{\bibfnamefont{T.}~\bibnamefont{Kanaya}},
  \bibinfo{journal}{J. Comp. Phys.} \textbf{\bibinfo{volume}{99}},
  \bibinfo{pages}{8200} (\bibinfo{year}{1993}).

\bibitem[{\citenamefont{Emerton et~al.}(1997)\citenamefont{Emerton, Coveney,
  and Boghosian}}]{bib:emerton-coveney-boghosian}
\bibinfo{author}{\bibfnamefont{A.~N.} \bibnamefont{Emerton}},
  \bibinfo{author}{\bibfnamefont{P.~V.} \bibnamefont{Coveney}},
  \bibnamefont{and} \bibinfo{author}{\bibfnamefont{B.~M.}
  \bibnamefont{Boghosian}}, \bibinfo{journal}{Phys. Rev. E}
  \textbf{\bibinfo{volume}{56}}, \bibinfo{pages}{1286} (\bibinfo{year}{1997}).

\bibitem[{\citenamefont{Gonz\'{a}lez-Segredo and
  Coveney}(2004{\natexlab{a}})}]{bib:gonzalez-coveney-2}
\bibinfo{author}{\bibfnamefont{N.}~\bibnamefont{Gonz\'{a}lez-Segredo}}
  \bibnamefont{and} \bibinfo{author}{\bibfnamefont{P.~V.}
  \bibnamefont{Coveney}}, \bibinfo{journal}{Phys. Rev. E}
  \textbf{\bibinfo{volume}{69}}, \bibinfo{pages}{061501}
  (\bibinfo{year}{2004}{\natexlab{a}}).

\bibitem[{\citenamefont{Jones}(2003)}]{bib:JonesBook}
\bibinfo{author}{\bibfnamefont{R.~A.~L.} \bibnamefont{Jones}},
  \emph{\bibinfo{title}{Soft Condensed Matter}} (\bibinfo{publisher}{Oxford
  University Press}, \bibinfo{year}{2003}).

\bibitem[{\citenamefont{Gompper and Schick}(1994)}]{bib:gompper-schick}
\bibinfo{author}{\bibfnamefont{G.}~\bibnamefont{Gompper}} \bibnamefont{and}
  \bibinfo{author}{\bibfnamefont{M.}~\bibnamefont{Schick}}, in
  \emph{\bibinfo{booktitle}{Phase Transitions and Critical Phenomena}}, edited
  by \bibinfo{editor}{\bibfnamefont{C.}~\bibnamefont{Domb}} \bibnamefont{and}
  \bibinfo{editor}{\bibfnamefont{J.}~\bibnamefont{Lebowitz}}
  (\bibinfo{publisher}{Academic Press}, \bibinfo{year}{1994}),
  vol.~\bibinfo{volume}{16}, pp. \bibinfo{pages}{1--176}.

\bibitem[{\citenamefont{Meyer et~al.}(2000)\citenamefont{Meyer, Asnacios,
  Bourgaux, and Kleman}}]{bib:meyer-2000}
\bibinfo{author}{\bibfnamefont{C.}~\bibnamefont{Meyer}},
  \bibinfo{author}{\bibfnamefont{S.}~\bibnamefont{Asnacios}},
  \bibinfo{author}{\bibfnamefont{C.}~\bibnamefont{Bourgaux}}, \bibnamefont{and}
  \bibinfo{author}{\bibfnamefont{M.}~\bibnamefont{Kleman}},
  \bibinfo{journal}{Rheol. Acta} \textbf{\bibinfo{volume}{39}},
  \bibinfo{pages}{223} (\bibinfo{year}{2000}).

\bibitem[{\citenamefont{Zipfel et~al.}(1999)\citenamefont{Zipfel, Berghausen,
  Schmidt, Lindner, Tsianou, Alexandridis, and Richtering}}]{bib:zipfel-1999}
\bibinfo{author}{\bibfnamefont{J.}~\bibnamefont{Zipfel}},
  \bibinfo{author}{\bibfnamefont{J.}~\bibnamefont{Berghausen}},
  \bibinfo{author}{\bibfnamefont{G.}~\bibnamefont{Schmidt}},
  \bibinfo{author}{\bibfnamefont{P.}~\bibnamefont{Lindner}},
  \bibinfo{author}{\bibfnamefont{M.}~\bibnamefont{Tsianou}},
  \bibinfo{author}{\bibfnamefont{P.}~\bibnamefont{Alexandridis}},
  \bibnamefont{and}
  \bibinfo{author}{\bibfnamefont{W.}~\bibnamefont{Richtering}},
  \bibinfo{journal}{Phys. Chem. Chem. Phys.} \textbf{\bibinfo{volume}{1}},
  \bibinfo{pages}{3905} (\bibinfo{year}{1999}).

\bibitem[{\citenamefont{Berghausen et~al.}(2000)\citenamefont{Berghausen,
  Zipfel, Diat, Narayanan, and Richtering}}]{bib:berghausen-2000}
\bibinfo{author}{\bibfnamefont{J.}~\bibnamefont{Berghausen}},
  \bibinfo{author}{\bibfnamefont{J.}~\bibnamefont{Zipfel}},
  \bibinfo{author}{\bibfnamefont{O.}~\bibnamefont{Diat}},
  \bibinfo{author}{\bibfnamefont{T.}~\bibnamefont{Narayanan}},
  \bibnamefont{and}
  \bibinfo{author}{\bibfnamefont{W.}~\bibnamefont{Richtering}},
  \bibinfo{journal}{Phys. Chem. Chem. Phys.} \textbf{\bibinfo{volume}{2}},
  \bibinfo{pages}{3623} (\bibinfo{year}{2000}).

\bibitem[{\citenamefont{Qiu et~al.}(1998)\citenamefont{Qiu, Zhang, and
  Yang}}]{bib:qiu-zhang-yang:1998}
\bibinfo{author}{\bibfnamefont{F.}~\bibnamefont{Qiu}},
  \bibinfo{author}{\bibfnamefont{H.}~\bibnamefont{Zhang}}, \bibnamefont{and}
  \bibinfo{author}{\bibfnamefont{Y.}~\bibnamefont{Yang}}, \bibinfo{journal}{J.
  Comp. Phys.} \textbf{\bibinfo{volume}{109}}, \bibinfo{pages}{1575}
  (\bibinfo{year}{1998}).

\bibitem[{\citenamefont{Zhang et~al.}(1996)\citenamefont{Zhang, Wiesner, Yang,
  Pakula, and Spiess}}]{bib:zhang-wiesner-yang-pakula-spiess:1996}
\bibinfo{author}{\bibfnamefont{Y.}~\bibnamefont{Zhang}},
  \bibinfo{author}{\bibfnamefont{U.}~\bibnamefont{Wiesner}},
  \bibinfo{author}{\bibfnamefont{Y.}~\bibnamefont{Yang}},
  \bibinfo{author}{\bibfnamefont{T.}~\bibnamefont{Pakula}}, \bibnamefont{and}
  \bibinfo{author}{\bibfnamefont{H.}~\bibnamefont{Spiess}},
  \bibinfo{journal}{Macromolecules} \textbf{\bibinfo{volume}{29}},
  \bibinfo{pages}{5427} (\bibinfo{year}{1996}).

\bibitem[{\citenamefont{Zhang and Wiesner}(1995)}]{bib:zhang-wiesner:1995}
\bibinfo{author}{\bibfnamefont{Y.}~\bibnamefont{Zhang}} \bibnamefont{and}
  \bibinfo{author}{\bibfnamefont{U.}~\bibnamefont{Wiesner}},
  \bibinfo{journal}{J. Comp. Phys.} \textbf{\bibinfo{volume}{103}},
  \bibinfo{pages}{4784} (\bibinfo{year}{1995}).

\bibitem[{\citenamefont{Xu et~al.}(2003)\citenamefont{Xu, Gonnella, and
  Lamura}}]{bib:xu-gonnella-lamura:2003}
\bibinfo{author}{\bibfnamefont{A.}~\bibnamefont{Xu}},
  \bibinfo{author}{\bibfnamefont{G.}~\bibnamefont{Gonnella}}, \bibnamefont{and}
  \bibinfo{author}{\bibfnamefont{A.}~\bibnamefont{Lamura}},
  \bibinfo{journal}{Phys. Rev. E} \textbf{\bibinfo{volume}{67}},
  \bibinfo{pages}{056105} (\bibinfo{year}{2003}).

\bibitem[{\citenamefont{Espa$\tilde{\mbox{n}}$ol and
  Warren}(1995)}]{bib:espanol-warren}
\bibinfo{author}{\bibfnamefont{P.}~\bibnamefont{Espa$\tilde{\mbox{n}}$ol}}
  \bibnamefont{and} \bibinfo{author}{\bibfnamefont{P.}~\bibnamefont{Warren}},
  \bibinfo{journal}{Europhys. Lett.} \textbf{\bibinfo{volume}{30}},
  \bibinfo{pages}{191} (\bibinfo{year}{1995}).

\bibitem[{\citenamefont{Jury et~al.}(1999)\citenamefont{Jury, Bladon, Cates,
  Krishna, Hagen, Ruddock, and Warren}}]{bib:jury-bladon-cates}
\bibinfo{author}{\bibfnamefont{S.}~\bibnamefont{Jury}},
  \bibinfo{author}{\bibfnamefont{P.}~\bibnamefont{Bladon}},
  \bibinfo{author}{\bibfnamefont{M.}~\bibnamefont{Cates}},
  \bibinfo{author}{\bibfnamefont{S.}~\bibnamefont{Krishna}},
  \bibinfo{author}{\bibfnamefont{M.}~\bibnamefont{Hagen}},
  \bibinfo{author}{\bibfnamefont{N.}~\bibnamefont{Ruddock}}, \bibnamefont{and}
  \bibinfo{author}{\bibfnamefont{P.}~\bibnamefont{Warren}},
  \bibinfo{journal}{Phys. Chem. Chem. Phys.} \textbf{\bibinfo{volume}{1}},
  \bibinfo{pages}{2051} (\bibinfo{year}{1999}).

\bibitem[{\citenamefont{Flekk{\o}y et~al.}(2000)\citenamefont{Flekk{\o}y,
  Coveney, and de~Fabritiis}}]{bib:flekkoy-coveney-defabritiis}
\bibinfo{author}{\bibfnamefont{E.~G.} \bibnamefont{Flekk{\o}y}},
  \bibinfo{author}{\bibfnamefont{P.~V.} \bibnamefont{Coveney}},
  \bibnamefont{and}
  \bibinfo{author}{\bibfnamefont{G.}~\bibnamefont{de~Fabritiis}},
  \bibinfo{journal}{Phys. Rev. E} \textbf{\bibinfo{volume}{62}},
  \bibinfo{pages}{2140} (\bibinfo{year}{2000}).

\bibitem[{\citenamefont{Rivet and Boon}(2001)}]{bib:rivet-boon}
\bibinfo{author}{\bibfnamefont{J.-P.} \bibnamefont{Rivet}} \bibnamefont{and}
  \bibinfo{author}{\bibfnamefont{J.~P.} \bibnamefont{Boon}},
  \emph{\bibinfo{title}{Lattice Gas Hydrodynamics}}
  (\bibinfo{publisher}{Cambridge University Press}, \bibinfo{year}{2001}).

\bibitem[{\citenamefont{Malevanets and Kapral}(1998)}]{bib:malevanets-kapral}
\bibinfo{author}{\bibfnamefont{A.}~\bibnamefont{Malevanets}} \bibnamefont{and}
  \bibinfo{author}{\bibfnamefont{R.}~\bibnamefont{Kapral}},
  \bibinfo{journal}{Europhys. Lett.} \textbf{\bibinfo{volume}{44}},
  \bibinfo{pages}{552} (\bibinfo{year}{1998}).

\bibitem[{\citenamefont{Hashimoto et~al.}(2000)\citenamefont{Hashimoto, Chen,
  and Ohashi}}]{bib:hashimoto-chen-ohashi}
\bibinfo{author}{\bibfnamefont{Y.}~\bibnamefont{Hashimoto}},
  \bibinfo{author}{\bibfnamefont{Y.}~\bibnamefont{Chen}}, \bibnamefont{and}
  \bibinfo{author}{\bibfnamefont{H.}~\bibnamefont{Ohashi}},
  \bibinfo{journal}{Comp. Phys. Comm.} \textbf{\bibinfo{volume}{129}},
  \bibinfo{pages}{56} (\bibinfo{year}{2000}).

\bibitem[{\citenamefont{Sakai et~al.}(2000)\citenamefont{Sakai, Chen, and
  Ohashi}}]{bib:sakai-chen-ohashi}
\bibinfo{author}{\bibfnamefont{T.}~\bibnamefont{Sakai}},
  \bibinfo{author}{\bibfnamefont{Y.}~\bibnamefont{Chen}}, \bibnamefont{and}
  \bibinfo{author}{\bibfnamefont{H.}~\bibnamefont{Ohashi}},
  \bibinfo{journal}{Comp. Phys. Comm.} \textbf{\bibinfo{volume}{129}},
  \bibinfo{pages}{75} (\bibinfo{year}{2000}).

\bibitem[{\citenamefont{Succi}(2001)}]{bib:succi}
\bibinfo{author}{\bibfnamefont{S.}~\bibnamefont{Succi}},
  \emph{\bibinfo{title}{The Lattice {B}oltzmann Equation for Fluid Dynamics and
  Beyond}} (\bibinfo{publisher}{Oxford University Press},
  \bibinfo{year}{2001}).

\bibitem[{\citenamefont{Benzi et~al.}(1992)\citenamefont{Benzi, Succi, and
  Vergassola}}]{bib:benzi-succi-vergassola}
\bibinfo{author}{\bibfnamefont{R.}~\bibnamefont{Benzi}},
  \bibinfo{author}{\bibfnamefont{S.}~\bibnamefont{Succi}}, \bibnamefont{and}
  \bibinfo{author}{\bibfnamefont{M.}~\bibnamefont{Vergassola}},
  \bibinfo{journal}{Phys. Rep.} \textbf{\bibinfo{volume}{222}},
  \bibinfo{pages}{145 } (\bibinfo{year}{1992}).

\bibitem[{\citenamefont{Love et~al.}(2003)\citenamefont{Love, Nekovee, Coveney,
  Chin, Gonz\'{a}lez-Segredo, and
  Martin}}]{bib:love-nekovee-coveney-chin-gonzalez-martin}
\bibinfo{author}{\bibfnamefont{P.~J.} \bibnamefont{Love}},
  \bibinfo{author}{\bibfnamefont{M.}~\bibnamefont{Nekovee}},
  \bibinfo{author}{\bibfnamefont{P.~V.} \bibnamefont{Coveney}},
  \bibinfo{author}{\bibfnamefont{J.}~\bibnamefont{Chin}},
  \bibinfo{author}{\bibfnamefont{N.}~\bibnamefont{Gonz\'{a}lez-Segredo}},
  \bibnamefont{and} \bibinfo{author}{\bibfnamefont{J.~M.~R.}
  \bibnamefont{Martin}}, \bibinfo{journal}{Comp. Phys. Comm.}
  \textbf{\bibinfo{volume}{153}}, \bibinfo{pages}{340} (\bibinfo{year}{2003}).

\bibitem[{\citenamefont{Chen et~al.}(2000)\citenamefont{Chen, Boghosian,
  Coveney, and Nekovee}}]{bib:chen-boghosian-coveney}
\bibinfo{author}{\bibfnamefont{H.}~\bibnamefont{Chen}},
  \bibinfo{author}{\bibfnamefont{B.~M.} \bibnamefont{Boghosian}},
  \bibinfo{author}{\bibfnamefont{P.~V.} \bibnamefont{Coveney}},
  \bibnamefont{and} \bibinfo{author}{\bibfnamefont{M.}~\bibnamefont{Nekovee}},
  \bibinfo{journal}{Proc. R. Soc. Lond. A} \textbf{\bibinfo{volume}{456}},
  \bibinfo{pages}{2043} (\bibinfo{year}{2000}).

\bibitem[{\citenamefont{Osborn et~al.}(1995)\citenamefont{Osborn, Orlandini,
  Swift, Yeomans, and Banavar}}]{bib:osborn-orlandini-swift-yeomans-banavar}
\bibinfo{author}{\bibfnamefont{W.~R.} \bibnamefont{Osborn}},
  \bibinfo{author}{\bibfnamefont{E.}~\bibnamefont{Orlandini}},
  \bibinfo{author}{\bibfnamefont{M.~R.} \bibnamefont{Swift}},
  \bibinfo{author}{\bibfnamefont{J.~M.} \bibnamefont{Yeomans}},
  \bibnamefont{and} \bibinfo{author}{\bibfnamefont{J.~R.}
  \bibnamefont{Banavar}}, \bibinfo{journal}{Phys. Rev. Lett.}
  \textbf{\bibinfo{volume}{75}}, \bibinfo{pages}{4031} (\bibinfo{year}{1995}).

\bibitem[{\citenamefont{Chin and Coveney}(2002)}]{bib:chin-coveney}
\bibinfo{author}{\bibfnamefont{J.}~\bibnamefont{Chin}} \bibnamefont{and}
  \bibinfo{author}{\bibfnamefont{P.~V.} \bibnamefont{Coveney}},
  \bibinfo{journal}{Phys. Rev. E} \textbf{\bibinfo{volume}{66}},
  \bibinfo{pages}{016303} (\bibinfo{year}{2002}).

\bibitem[{\citenamefont{Chin et~al.}(2002)\citenamefont{Chin, Boek, and
  Coveney}}]{bib:chin-boek-coveney}
\bibinfo{author}{\bibfnamefont{J.}~\bibnamefont{Chin}},
  \bibinfo{author}{\bibfnamefont{E.~S.} \bibnamefont{Boek}}, \bibnamefont{and}
  \bibinfo{author}{\bibfnamefont{P.~V.} \bibnamefont{Coveney}},
  \bibinfo{journal}{Proc. R. Soc. Lond. A} \textbf{\bibinfo{volume}{360}},
  \bibinfo{pages}{547} (\bibinfo{year}{2002}).

\bibitem[{\citenamefont{Alexander et~al.}(1993)\citenamefont{Alexander, Chen,
  and Grunau}}]{bib:alexander-chen-grunau}
\bibinfo{author}{\bibfnamefont{F.~J.} \bibnamefont{Alexander}},
  \bibinfo{author}{\bibfnamefont{S.}~\bibnamefont{Chen}}, \bibnamefont{and}
  \bibinfo{author}{\bibfnamefont{D.~W.} \bibnamefont{Grunau}},
  \bibinfo{journal}{Phys. Rev. B} \textbf{\bibinfo{volume}{48}},
  \bibinfo{pages}{634} (\bibinfo{year}{1993}).

\bibitem[{\citenamefont{Wagner and Yeomans}(1998)}]{bib:wagner-yeomans}
\bibinfo{author}{\bibfnamefont{A.~J.} \bibnamefont{Wagner}} \bibnamefont{and}
  \bibinfo{author}{\bibfnamefont{J.~M.} \bibnamefont{Yeomans}},
  \bibinfo{journal}{Phys. Rev. Lett.} \textbf{\bibinfo{volume}{80}},
  \bibinfo{pages}{1429} (\bibinfo{year}{1998}).

\bibitem[{\citenamefont{Gonnella et~al.}(1997)\citenamefont{Gonnella,
  Orlandini, and Yeomans}}]{bib:gonnella-orlandini-yeomans}
\bibinfo{author}{\bibfnamefont{G.}~\bibnamefont{Gonnella}},
  \bibinfo{author}{\bibfnamefont{E.}~\bibnamefont{Orlandini}},
  \bibnamefont{and} \bibinfo{author}{\bibfnamefont{J.~M.}
  \bibnamefont{Yeomans}}, \bibinfo{journal}{Phys. Rev. Lett.}
  \textbf{\bibinfo{volume}{78}}, \bibinfo{pages}{1695} (\bibinfo{year}{1997}).

\bibitem[{\citenamefont{Cates et~al.}(1999)\citenamefont{Cates, Kendon, Bladon,
  and Desplat}}]{bib:cates-kendon-bladon-desplat}
\bibinfo{author}{\bibfnamefont{M.~E.} \bibnamefont{Cates}},
  \bibinfo{author}{\bibfnamefont{V.~M.} \bibnamefont{Kendon}},
  \bibinfo{author}{\bibfnamefont{P.}~\bibnamefont{Bladon}}, \bibnamefont{and}
  \bibinfo{author}{\bibfnamefont{J.~C.} \bibnamefont{Desplat}},
  \bibinfo{journal}{Faraday Disc.} \textbf{\bibinfo{volume}{112}},
  \bibinfo{pages}{1} (\bibinfo{year}{1999}).

\bibitem[{\citenamefont{Kendon et~al.}(1999)\citenamefont{Kendon, Desplat,
  Bladon, and Cates}}]{bib:kendon-desplat-bladon-cates}
\bibinfo{author}{\bibfnamefont{V.~M.} \bibnamefont{Kendon}},
  \bibinfo{author}{\bibfnamefont{J.~C.} \bibnamefont{Desplat}},
  \bibinfo{author}{\bibfnamefont{P.}~\bibnamefont{Bladon}}, \bibnamefont{and}
  \bibinfo{author}{\bibfnamefont{M.~E.} \bibnamefont{Cates}},
  \bibinfo{journal}{Phys. Rev. Lett.} \textbf{\bibinfo{volume}{83}},
  \bibinfo{pages}{576} (\bibinfo{year}{1999}).

\bibitem[{\citenamefont{Kendon et~al.}(2001)\citenamefont{Kendon, Cates,
  Pagonabarraga, Desplat, and
  Bladon}}]{bib:kendon-cates-pagonabarraga-desplat-bladon}
\bibinfo{author}{\bibfnamefont{V.~M.} \bibnamefont{Kendon}},
  \bibinfo{author}{\bibfnamefont{M.~E.} \bibnamefont{Cates}},
  \bibinfo{author}{\bibfnamefont{I.}~\bibnamefont{Pagonabarraga}},
  \bibinfo{author}{\bibfnamefont{J.~C.} \bibnamefont{Desplat}},
  \bibnamefont{and} \bibinfo{author}{\bibfnamefont{P.}~\bibnamefont{Bladon}},
  \bibinfo{journal}{J. Fluid Mech.} \textbf{\bibinfo{volume}{440}},
  \bibinfo{pages}{147} (\bibinfo{year}{2001}).

\bibitem[{\citenamefont{Pagonabarraga et~al.}(2001)\citenamefont{Pagonabarraga,
  Desplat, Wagner, and Cates}}]{bib:pagonabarraga-desplat-wagner-cates:2001}
\bibinfo{author}{\bibfnamefont{I.}~\bibnamefont{Pagonabarraga}},
  \bibinfo{author}{\bibfnamefont{J.~C.} \bibnamefont{Desplat}},
  \bibinfo{author}{\bibfnamefont{A.~J.} \bibnamefont{Wagner}},
  \bibnamefont{and} \bibinfo{author}{\bibfnamefont{M.~E.} \bibnamefont{Cates}},
  \bibinfo{journal}{New J. Phys.} \textbf{\bibinfo{volume}{3}},
  \bibinfo{pages}{9.1} (\bibinfo{year}{2001}).

\bibitem[{\citenamefont{Xu et~al.}(2004)\citenamefont{Xu, Gonnella, and
  Lamura}}]{bib:xu-gonnella-lamura:2004}
\bibinfo{author}{\bibfnamefont{A.}~\bibnamefont{Xu}},
  \bibinfo{author}{\bibfnamefont{G.}~\bibnamefont{Gonnella}}, \bibnamefont{and}
  \bibinfo{author}{\bibfnamefont{A.}~\bibnamefont{Lamura}},
  \bibinfo{journal}{Physica A} \textbf{\bibinfo{volume}{331}},
  \bibinfo{pages}{10} (\bibinfo{year}{2004}).

\bibitem[{\citenamefont{Wagner and Yeomans}(1999)}]{bib:wagner-yeomans-shear}
\bibinfo{author}{\bibfnamefont{A.~J.} \bibnamefont{Wagner}} \bibnamefont{and}
  \bibinfo{author}{\bibfnamefont{J.~M.} \bibnamefont{Yeomans}},
  \bibinfo{journal}{Phys. Rev. E} \textbf{\bibinfo{volume}{59}},
  \bibinfo{pages}{4366} (\bibinfo{year}{1999}).

\bibitem[{\citenamefont{Wagner and
  Pagonabarraga}(2002)}]{bib:wagner-pagonabarraga}
\bibinfo{author}{\bibfnamefont{A.}~\bibnamefont{Wagner}} \bibnamefont{and}
  \bibinfo{author}{\bibfnamefont{I.}~\bibnamefont{Pagonabarraga}},
  \bibinfo{journal}{J. Stat. Phys.} \textbf{\bibinfo{volume}{107}},
  \bibinfo{pages}{521} (\bibinfo{year}{2002}).

\bibitem[{\citenamefont{Harting
  et~al.}(2004{\natexlab{a}})\citenamefont{Harting, Venturoli, and
  Coveney}}]{bib:jens-venturoli-coveney:2004}
\bibinfo{author}{\bibfnamefont{J.}~\bibnamefont{Harting}},
  \bibinfo{author}{\bibfnamefont{M.}~\bibnamefont{Venturoli}},
  \bibnamefont{and} \bibinfo{author}{\bibfnamefont{P.~V.}
  \bibnamefont{Coveney}}, \bibinfo{journal}{Phil. Trans. R. Soc. Lond. A}
  \textbf{\bibinfo{volume}{362}}, \bibinfo{pages}{1703}
  (\bibinfo{year}{2004}{\natexlab{a}}).

\bibitem[{\citenamefont{Stansell et~al.}(2006)\citenamefont{Stansell,
  Stratford, Desplat, Adhikari, and
  Cates}}]{bib:stansell-stratford-desplat-adhikari-cates:2006}
\bibinfo{author}{\bibfnamefont{P.}~\bibnamefont{Stansell}},
  \bibinfo{author}{\bibfnamefont{K.}~\bibnamefont{Stratford}},
  \bibinfo{author}{\bibfnamefont{J.~C.} \bibnamefont{Desplat}},
  \bibinfo{author}{\bibfnamefont{R.}~\bibnamefont{Adhikari}}, \bibnamefont{and}
  \bibinfo{author}{\bibfnamefont{M.~E.} \bibnamefont{Cates}},
  \bibinfo{journal}{Phys. Rev. Lett.} \textbf{\bibinfo{volume}{96}},
  \bibinfo{pages}{085701} (\bibinfo{year}{2006}).

\bibitem[{\citenamefont{Theissen et~al.}(1998)\citenamefont{Theissen, Gompper,
  and Kroll}}]{bib:theissen-gompper-kroll}
\bibinfo{author}{\bibfnamefont{O.}~\bibnamefont{Theissen}},
  \bibinfo{author}{\bibfnamefont{G.}~\bibnamefont{Gompper}}, \bibnamefont{and}
  \bibinfo{author}{\bibfnamefont{D.~M.} \bibnamefont{Kroll}},
  \bibinfo{journal}{Europhys. Lett.} \textbf{\bibinfo{volume}{42}},
  \bibinfo{pages}{419} (\bibinfo{year}{1998}).

\bibitem[{\citenamefont{Lamura et~al.}(1999)\citenamefont{Lamura, Gonnella, and
  Yeomans}}]{bib:lamura-gonnella-yeomans}
\bibinfo{author}{\bibfnamefont{A.}~\bibnamefont{Lamura}},
  \bibinfo{author}{\bibfnamefont{G.}~\bibnamefont{Gonnella}}, \bibnamefont{and}
  \bibinfo{author}{\bibfnamefont{J.~M.} \bibnamefont{Yeomans}},
  \bibinfo{journal}{Europhys. Lett.} \textbf{\bibinfo{volume}{45}},
  \bibinfo{pages}{314} (\bibinfo{year}{1999}).

\bibitem[{\citenamefont{Nekovee et~al.}(2000)\citenamefont{Nekovee, Coveney,
  Chen, and Boghosian}}]{bib:nekovee-coveney-chen-boghosian}
\bibinfo{author}{\bibfnamefont{M.}~\bibnamefont{Nekovee}},
  \bibinfo{author}{\bibfnamefont{P.~V.} \bibnamefont{Coveney}},
  \bibinfo{author}{\bibfnamefont{H.}~\bibnamefont{Chen}}, \bibnamefont{and}
  \bibinfo{author}{\bibfnamefont{B.~M.} \bibnamefont{Boghosian}},
  \bibinfo{journal}{Phys. Rev. E} \textbf{\bibinfo{volume}{62}},
  \bibinfo{pages}{8282} (\bibinfo{year}{2000}).

\bibitem[{\citenamefont{Nekovee and
  Coveney}(2001{\natexlab{a}})}]{bib:nekovee-coveney}
\bibinfo{author}{\bibfnamefont{M.}~\bibnamefont{Nekovee}} \bibnamefont{and}
  \bibinfo{author}{\bibfnamefont{P.~V.} \bibnamefont{Coveney}},
  \bibinfo{journal}{J. Am. Chem. Soc.} \textbf{\bibinfo{volume}{123}},
  \bibinfo{pages}{12380} (\bibinfo{year}{2001}{\natexlab{a}}).

\bibitem[{\citenamefont{Giupponi et~al.}(2006)\citenamefont{Giupponi, Harting,
  and Coveney}}]{bib:jens-giupponi-coveney:2006}
\bibinfo{author}{\bibfnamefont{G.}~\bibnamefont{Giupponi}},
  \bibinfo{author}{\bibfnamefont{J.}~\bibnamefont{Harting}}, \bibnamefont{and}
  \bibinfo{author}{\bibfnamefont{P.~V.} \bibnamefont{Coveney}},
  \bibinfo{journal}{Europhys. Lett.} \textbf{\bibinfo{volume}{73}},
  \bibinfo{pages}{533} (\bibinfo{year}{2006}).

\bibitem[{\citenamefont{Gonz\'{a}lez-Segredo and
  Coveney}(2004{\natexlab{b}})}]{bib:gonzalez-coveney}
\bibinfo{author}{\bibfnamefont{N.}~\bibnamefont{Gonz\'{a}lez-Segredo}}
  \bibnamefont{and} \bibinfo{author}{\bibfnamefont{P.~V.}
  \bibnamefont{Coveney}}, \bibinfo{journal}{Europhys. Lett.}
  \textbf{\bibinfo{volume}{65}}, \bibinfo{pages}{795}
  (\bibinfo{year}{2004}{\natexlab{b}}).

\bibitem[{\citenamefont{Pickles et~al.}(2004)\citenamefont{Pickles, Blake,
  Boghosian, Brooke, Chin, Clarke, Coveney, Haines, Harting, Harvey
  et~al.}}]{bib:jens-teragyroid:2004}
\bibinfo{author}{\bibfnamefont{S.~M.} \bibnamefont{Pickles}},
  \bibinfo{author}{\bibfnamefont{R.~J.} \bibnamefont{Blake}},
  \bibinfo{author}{\bibfnamefont{B.~M.} \bibnamefont{Boghosian}},
  \bibinfo{author}{\bibfnamefont{J.~M.} \bibnamefont{Brooke}},
  \bibinfo{author}{\bibfnamefont{J.}~\bibnamefont{Chin}},
  \bibinfo{author}{\bibfnamefont{P.~E.~L.} \bibnamefont{Clarke}},
  \bibinfo{author}{\bibfnamefont{P.~V.} \bibnamefont{Coveney}},
  \bibinfo{author}{\bibfnamefont{R.}~\bibnamefont{Haines}},
  \bibinfo{author}{\bibfnamefont{J.}~\bibnamefont{Harting}},
  \bibinfo{author}{\bibfnamefont{M.}~\bibnamefont{Harvey}},
  \bibnamefont{et~al.}, \bibinfo{journal}{Proceedings of the Workshop on Case
  Studies on Grid Applications at GGF 10}  (\bibinfo{year}{2004}),
  \bibinfo{note}{available online:
  http://www.zib.de/ggf/apps/meetings/ggf10.html}.

\bibitem[{\citenamefont{Harting
  et~al.}(2004{\natexlab{b}})\citenamefont{Harting, Harvey, Chin, and
  Coveney}}]{bib:jens-harvey-chin-coveney:2004}
\bibinfo{author}{\bibfnamefont{J.}~\bibnamefont{Harting}},
  \bibinfo{author}{\bibfnamefont{M.}~\bibnamefont{Harvey}},
  \bibinfo{author}{\bibfnamefont{J.}~\bibnamefont{Chin}}, \bibnamefont{and}
  \bibinfo{author}{\bibfnamefont{P.~V.} \bibnamefont{Coveney}},
  \bibinfo{journal}{Comp. Phys. Comm.} \textbf{\bibinfo{volume}{165}},
  \bibinfo{pages}{97} (\bibinfo{year}{2004}{\natexlab{b}}).

\bibitem[{\citenamefont{Harting et~al.}(2005)\citenamefont{Harting, Harvey,
  Chin, Venturoli, and Coveney}}]{bib:jens-harvey-chin-venturoli-coveney:2005}
\bibinfo{author}{\bibfnamefont{J.}~\bibnamefont{Harting}},
  \bibinfo{author}{\bibfnamefont{M.}~\bibnamefont{Harvey}},
  \bibinfo{author}{\bibfnamefont{J.}~\bibnamefont{Chin}},
  \bibinfo{author}{\bibfnamefont{M.}~\bibnamefont{Venturoli}},
  \bibnamefont{and} \bibinfo{author}{\bibfnamefont{P.~V.}
  \bibnamefont{Coveney}}, \bibinfo{journal}{Phil. Trans. R. Soc. Lond. A}
  \textbf{\bibinfo{volume}{363}}, \bibinfo{pages}{1895} (\bibinfo{year}{2005}).

\bibitem[{\citenamefont{Chin and Coveney}(2006)}]{bib:chin-coveney06}
\bibinfo{author}{\bibfnamefont{J.}~\bibnamefont{Chin}} \bibnamefont{and}
  \bibinfo{author}{\bibfnamefont{P.~V.} \bibnamefont{Coveney}},
  \bibinfo{journal}{Phil. Trans. R. Soc. Lond. A}
  \textbf{\bibinfo{volume}{462}} (\bibinfo{year}{2006}).

\bibitem[{\citenamefont{Gonz\'{a}lez-Segredo
  et~al.}(2006)\citenamefont{Gonz\'{a}lez-Segredo, Harting, Giupponi, and
  Coveney}}]{bib:jens-gonzalez-giupponi-coveney:2005}
\bibinfo{author}{\bibfnamefont{N.}~\bibnamefont{Gonz\'{a}lez-Segredo}},
  \bibinfo{author}{\bibfnamefont{J.}~\bibnamefont{Harting}},
  \bibinfo{author}{\bibfnamefont{G.}~\bibnamefont{Giupponi}}, \bibnamefont{and}
  \bibinfo{author}{\bibfnamefont{P.~V.} \bibnamefont{Coveney}},
  \bibinfo{journal}{Phys. Rev. E} \textbf{\bibinfo{volume}{73}},
  \bibinfo{pages}{031503} (\bibinfo{year}{2006}).

\bibitem[{\citenamefont{Higuera et~al.}(1989)\citenamefont{Higuera, Succi, and
  Benzi}}]{bib:higuera-succi-benzi}
\bibinfo{author}{\bibfnamefont{P.~J.} \bibnamefont{Higuera}},
  \bibinfo{author}{\bibfnamefont{S.}~\bibnamefont{Succi}}, \bibnamefont{and}
  \bibinfo{author}{\bibfnamefont{R.}~\bibnamefont{Benzi}},
  \bibinfo{journal}{Europhys. Lett.} \textbf{\bibinfo{volume}{9}},
  \bibinfo{pages}{345} (\bibinfo{year}{1989}).

\bibitem[{\citenamefont{Bhatnagar et~al.}(1954)\citenamefont{Bhatnagar, Gross,
  and Krook}}]{bib:Bhatnagar54}
\bibinfo{author}{\bibfnamefont{P.~L.} \bibnamefont{Bhatnagar}},
  \bibinfo{author}{\bibfnamefont{E.~P.} \bibnamefont{Gross}}, \bibnamefont{and}
  \bibinfo{author}{\bibfnamefont{M.}~\bibnamefont{Krook}},
  \bibinfo{journal}{Phys. Rev.} \textbf{\bibinfo{volume}{94}},
  \bibinfo{pages}{511} (\bibinfo{year}{1954}).

\bibitem[{\citenamefont{Chen et~al.}(1991)\citenamefont{Chen, Chen,
  Mart\'{\i}nez, and Matthaeus}}]{bib:chen-chen-martinez-matthaeus}
\bibinfo{author}{\bibfnamefont{S.}~\bibnamefont{Chen}},
  \bibinfo{author}{\bibfnamefont{H.}~\bibnamefont{Chen}},
  \bibinfo{author}{\bibfnamefont{D.}~\bibnamefont{Mart\'{\i}nez}},
  \bibnamefont{and}
  \bibinfo{author}{\bibfnamefont{W.}~\bibnamefont{Matthaeus}},
  \bibinfo{journal}{Phys. Rev. Lett.} \textbf{\bibinfo{volume}{67}},
  \bibinfo{pages}{3776} (\bibinfo{year}{1991}).

\bibitem[{\citenamefont{Shan and Chen}(1993)}]{bib:shan-chen}
\bibinfo{author}{\bibfnamefont{X.}~\bibnamefont{Shan}} \bibnamefont{and}
  \bibinfo{author}{\bibfnamefont{H.}~\bibnamefont{Chen}},
  \bibinfo{journal}{Phys. Rev. E} \textbf{\bibinfo{volume}{47}},
  \bibinfo{pages}{1815} (\bibinfo{year}{1993}).

\bibitem[{\citenamefont{Shan and Chen}(1994)}]{bib:shan-chen-liq-gas}
\bibinfo{author}{\bibfnamefont{X.}~\bibnamefont{Shan}} \bibnamefont{and}
  \bibinfo{author}{\bibfnamefont{H.}~\bibnamefont{Chen}},
  \bibinfo{journal}{Phys. Rev. E} \textbf{\bibinfo{volume}{49}},
  \bibinfo{pages}{2941} (\bibinfo{year}{1994}).

\bibitem[{\citenamefont{Nekovee and Coveney}(2001{\natexlab{b}})}]{Maziar:2001}
\bibinfo{author}{\bibfnamefont{M.}~\bibnamefont{Nekovee}} \bibnamefont{and}
  \bibinfo{author}{\bibfnamefont{P.~V.} \bibnamefont{Coveney}},
  \bibinfo{journal}{J.~Am.~Chem.~Soc.} \textbf{\bibinfo{volume}{123}},
  \bibinfo{pages}{12380} (\bibinfo{year}{2001}{\natexlab{b}}).

\bibitem[{\citenamefont{Lees and Edwards}(1972)}]{bib:lees-edwards}
\bibinfo{author}{\bibfnamefont{A.}~\bibnamefont{Lees}} \bibnamefont{and}
  \bibinfo{author}{\bibfnamefont{S.}~\bibnamefont{Edwards}},
  \bibinfo{journal}{J. Phys. C.} \textbf{\bibinfo{volume}{5}},
  \bibinfo{pages}{1921} (\bibinfo{year}{1972}).

\bibitem[{\citenamefont{Harting
  et~al.}(2004{\natexlab{c}})\citenamefont{Harting, Venturoli, and
  Coveney}}]{bib:harting-venturoli-coveney}
\bibinfo{author}{\bibfnamefont{J.}~\bibnamefont{Harting}},
  \bibinfo{author}{\bibfnamefont{M.}~\bibnamefont{Venturoli}},
  \bibnamefont{and} \bibinfo{author}{\bibfnamefont{P.~V.}
  \bibnamefont{Coveney}}, \bibinfo{journal}{Phil. Trans. R. Soc. Lond. A}
  \textbf{\bibinfo{volume}{362}}, \bibinfo{pages}{1703}
  (\bibinfo{year}{2004}{\natexlab{c}}).

\bibitem[{\citenamefont{Love et~al.}(2001)\citenamefont{Love, Coveney, and
  Boghosian}}]{bib:love-coveney-boghosian-2001}
\bibinfo{author}{\bibfnamefont{P.~J.} \bibnamefont{Love}},
  \bibinfo{author}{\bibfnamefont{P.~V.} \bibnamefont{Coveney}},
  \bibnamefont{and} \bibinfo{author}{\bibfnamefont{B.~M.}
  \bibnamefont{Boghosian}}, \bibinfo{journal}{Phys. Rev. E}
  \textbf{\bibinfo{volume}{64}}, \bibinfo{pages}{021503}
  (\bibinfo{year}{2001}).

\bibitem[{\citenamefont{Rothman}(1991)}]{bib:rothman:1991}
\bibinfo{author}{\bibfnamefont{D.}~\bibnamefont{Rothman}},
  \bibinfo{journal}{Europhys. Lett.} \textbf{\bibinfo{volume}{14}},
  \bibinfo{pages}{337} (\bibinfo{year}{1991}).

\bibitem[{\citenamefont{Krall et~al.}(1992)\citenamefont{Krall, Sengers, and
  Hamano}}]{bib:krall-sengers-hamano:1992}
\bibinfo{author}{\bibfnamefont{A.~H.} \bibnamefont{Krall}},
  \bibinfo{author}{\bibfnamefont{J.~V.} \bibnamefont{Sengers}},
  \bibnamefont{and} \bibinfo{author}{\bibfnamefont{K.}~\bibnamefont{Hamano}},
  \bibinfo{journal}{Phys. Rev. Lett.} \textbf{\bibinfo{volume}{69}},
  \bibinfo{pages}{1963} (\bibinfo{year}{1992}).

\bibitem[{\citenamefont{Ohta et~al.}(1990)\citenamefont{Ohta, Nozaki, and
  Doi}}]{bib:ohta-nozaki-doi:1990}
\bibinfo{author}{\bibfnamefont{T.}~\bibnamefont{Ohta}},
  \bibinfo{author}{\bibfnamefont{H.}~\bibnamefont{Nozaki}}, \bibnamefont{and}
  \bibinfo{author}{\bibfnamefont{M.}~\bibnamefont{Doi}}, \bibinfo{journal}{J.
  Comp. Phys.} \textbf{\bibinfo{volume}{93}}, \bibinfo{pages}{2664}
  (\bibinfo{year}{1990}).

\bibitem[{\citenamefont{Corberi et~al.}(2000)\citenamefont{Corberi, Gonnella,
  and Lamura}}]{bib:corberi-gonnella-lamura}
\bibinfo{author}{\bibfnamefont{F.}~\bibnamefont{Corberi}},
  \bibinfo{author}{\bibfnamefont{G.}~\bibnamefont{Gonnella}}, \bibnamefont{and}
  \bibinfo{author}{\bibfnamefont{A.}~\bibnamefont{Lamura}},
  \bibinfo{journal}{Phys. Rev. E} \textbf{\bibinfo{volume}{62}},
  \bibinfo{pages}{8064} (\bibinfo{year}{2000}).

\bibitem[{\citenamefont{Hashimoto et~al.}(1994)\citenamefont{Hashimoto,
  Matsuzaka, Moses, and Onuki}}]{bib:hashimoto-matsuzaka-moses-onuki-1994}
\bibinfo{author}{\bibfnamefont{T.}~\bibnamefont{Hashimoto}},
  \bibinfo{author}{\bibfnamefont{K.}~\bibnamefont{Matsuzaka}},
  \bibinfo{author}{\bibfnamefont{E.}~\bibnamefont{Moses}}, \bibnamefont{and}
  \bibinfo{author}{\bibfnamefont{A.}~\bibnamefont{Onuki}},
  \bibinfo{journal}{Phys. Rev. Lett.} \textbf{\bibinfo{volume}{74}},
  \bibinfo{pages}{126} (\bibinfo{year}{1994}).

\bibitem[{\citenamefont{Shou and
  Chakrabarti}(2000)}]{bib:shou-chakrabarti:2000}
\bibinfo{author}{\bibfnamefont{Z.}~\bibnamefont{Shou}} \bibnamefont{and}
  \bibinfo{author}{\bibfnamefont{A.}~\bibnamefont{Chakrabarti}},
  \bibinfo{journal}{Phys. Rev. E} \textbf{\bibinfo{volume}{61}},
  \bibinfo{pages}{R2200} (\bibinfo{year}{2000}).

\bibitem[{\citenamefont{Corberi et~al.}(1998)\citenamefont{Corberi, Gonnella,
  and Lamura}}]{bib:corberi-gonnella-lamura:1998}
\bibinfo{author}{\bibfnamefont{F.}~\bibnamefont{Corberi}},
  \bibinfo{author}{\bibfnamefont{G.}~\bibnamefont{Gonnella}}, \bibnamefont{and}
  \bibinfo{author}{\bibfnamefont{A.}~\bibnamefont{Lamura}},
  \bibinfo{journal}{Phys. Rev. Lett.} \textbf{\bibinfo{volume}{81}},
  \bibinfo{pages}{3852} (\bibinfo{year}{1998}).

\bibitem[{\citenamefont{Corberi et~al.}(1999)\citenamefont{Corberi, Gonnella,
  and Lamura}}]{bib:corberi-gonnella-lamura:1999}
\bibinfo{author}{\bibfnamefont{F.}~\bibnamefont{Corberi}},
  \bibinfo{author}{\bibfnamefont{G.}~\bibnamefont{Gonnella}}, \bibnamefont{and}
  \bibinfo{author}{\bibfnamefont{A.}~\bibnamefont{Lamura}},
  \bibinfo{journal}{Phys. Rev. Lett.} \textbf{\bibinfo{volume}{83}},
  \bibinfo{pages}{4057} (\bibinfo{year}{1999}).

\bibitem[{\citenamefont{Corberi et~al.}(2002)\citenamefont{Corberi, Gonnella,
  and Lamura}}]{bib:corberi-gonnella-lamura:2002}
\bibinfo{author}{\bibfnamefont{F.}~\bibnamefont{Corberi}},
  \bibinfo{author}{\bibfnamefont{G.}~\bibnamefont{Gonnella}}, \bibnamefont{and}
  \bibinfo{author}{\bibfnamefont{A.}~\bibnamefont{Lamura}},
  \bibinfo{journal}{Phys. Rev. E} \textbf{\bibinfo{volume}{66}},
  \bibinfo{pages}{016114} (\bibinfo{year}{2002}).

\bibitem[{\citenamefont{Corberi et~al.}(2001)\citenamefont{Corberi, Gonnella,
  and Suppa}}]{bib:corberi-gonnella-suppa:2001}
\bibinfo{author}{\bibfnamefont{F.}~\bibnamefont{Corberi}},
  \bibinfo{author}{\bibfnamefont{G.}~\bibnamefont{Gonnella}}, \bibnamefont{and}
  \bibinfo{author}{\bibfnamefont{D.}~\bibnamefont{Suppa}},
  \bibinfo{journal}{Phys. Rev. E} \textbf{\bibinfo{volume}{63}},
  \bibinfo{pages}{040501(R)} (\bibinfo{year}{2001}).

\bibitem[{\citenamefont{Krall et~al.}(1993)\citenamefont{Krall, Sengers, and
  Hamano}}]{bib:krall-sengers-hamano:1993}
\bibinfo{author}{\bibfnamefont{A.~H.} \bibnamefont{Krall}},
  \bibinfo{author}{\bibfnamefont{J.~V.} \bibnamefont{Sengers}},
  \bibnamefont{and} \bibinfo{author}{\bibfnamefont{K.}~\bibnamefont{Hamano}},
  \bibinfo{journal}{Phys. Rev. E} \textbf{\bibinfo{volume}{48}},
  \bibinfo{pages}{357} (\bibinfo{year}{1993}).

\bibitem[{\citenamefont{Matsuzaka et~al.}(1997)\citenamefont{Matsuzaka, Jinnai,
  Koga, and Hashimoto}}]{bib:matsuzaka-jinnai-etal:1997}
\bibinfo{author}{\bibfnamefont{K.}~\bibnamefont{Matsuzaka}},
  \bibinfo{author}{\bibfnamefont{H.}~\bibnamefont{Jinnai}},
  \bibinfo{author}{\bibfnamefont{T.}~\bibnamefont{Koga}}, \bibnamefont{and}
  \bibinfo{author}{\bibfnamefont{T.}~\bibnamefont{Hashimoto}},
  \bibinfo{journal}{Macromolecules} \textbf{\bibinfo{volume}{30}},
  \bibinfo{pages}{1146} (\bibinfo{year}{1997}).

\end{thebibliography}

\end{document}